\documentclass[aps, prx, reprint,footinbib,floatfix,superscriptaddress]{revtex4-2}
\usepackage{lipsum}
\usepackage{graphicx}
\usepackage{indentfirst}
\usepackage{braket}
\usepackage{float}
\usepackage{amsmath}
\usepackage{physics}
\usepackage{amssymb}
\usepackage{CJK}
\usepackage{esint}
\usepackage{color}
\usepackage[T1]{fontenc}
\usepackage{amsfonts}
\usepackage{footmisc}
\usepackage{scrextend}
\usepackage{multirow}
\usepackage[outdir=./]{epstopdf}
\usepackage{xcolor}
\usepackage[english]{babel}
\usepackage{url}
\usepackage{comment}
\usepackage{array}
\usepackage{subfiles}
\usepackage{tikz}
\usetikzlibrary{shapes,arrows}
\usepackage{mathrsfs}
\usepackage{mathtools}
\usepackage[mathscr]{eucal}

\usepackage[hyperfootnotes=false]{hyperref}
\usepackage{cleveref}
\definecolor{darkblue}{rgb}{0,0,0.5}
\hypersetup{
colorlinks=true,
linkcolor=black,
filecolor=blue,
citecolor=darkblue,  
urlcolor=cyan,
}

\usepackage{trimclip}

\makeatletter
\DeclareRobustCommand{\shortto}{%
  \mathrel{\mathpalette\short@to\relax}%
}

\newcommand{\short@to}[2]{%
  \mkern2mu
  \clipbox{{.5\width} 0 0 0}{$\m@th#1\vphantom{+}{\shortrightarrow}$}%
  }
\makeatother

\usepackage{pict2e,picture,graphicx}

\makeatletter
\DeclareRobustCommand{\Arrow}[1][]{%
\check@mathfonts
\if\relax\detokenize{#1}\relax
\settowidth{\dimen@}{$\m@th\rightarrow$}%
\else
\setlength{\dimen@}{#1}%
\fi
\sbox\z@{\usefont{U}{lasy}{m}{n}\symbol{41}}%
\begin{picture}(\dimen@,\ht\z@)
\roundcap
\put(\dimexpr\dimen@-.7\wd\z@,0){\usebox\z@}
\put(0,\fontdimen22\textfont2){\line(1,0){\dimen@}}
\end{picture}%
}
\makeatother

\def\be{\begin{equation}}
\def\ee{\end{equation}}
\def\ba{\begin{eqnarray}}
\def\ea{\end{eqnarray}}
\def\bal{\begin{equation}\begin{aligned}}
\def\eal{\end{aligned}\end{equation}}

\def\bp{\begin{pmatrix}}
\def\ep{\end{pmatrix}}

\usepackage{bm}

\newcommand{\calE}{{\cal E}}

\newcommand{\calN}{{\cal N}}

\newcommand{\1}{^{(1)}}

\newcommand{\QZ}[1]{{{\textcolor{black}{#1}}}}

\usepackage[normalem]{ulem}

\begin{document}

\title{Entanglement-enhanced dual-comb spectroscopy}
\author{Haowei Shi}
\affiliation{
Ming Hsieh Department of Electrical and Computer Engineering, University of Southern California, Los
Angeles, California 90089, USA
}

\author{Zaijun Chen}
\affiliation{
Ming Hsieh Department of Electrical and Computer Engineering, University of Southern California, Los
Angeles, California 90089, USA
}

\author{Scott E. Fraser}
\affiliation{
Translational Imaging Center, University of Southern California, Los Angeles, California 90089, USA
}

\author{Mengjie Yu}
\affiliation{
Ming Hsieh Department of Electrical and Computer Engineering, University of Southern California, Los
Angeles, California 90089, USA
}

\author{Zheshen Zhang}
\affiliation{
Department of Electrical Engineering and Computer Science,
University of Michigan, Ann Arbor, MI 48109, USA
}

\author{Quntao Zhuang}
\email{qzhuang@usc.edu}
\affiliation{
Ming Hsieh Department of Electrical and Computer Engineering, University of Southern California, Los
Angeles, California 90089, USA
}
\affiliation{
Department of Physics and Astronomy, University of Southern California, Los
Angeles, California 90089, USA
}

\begin{abstract}
 Dual-comb interferometry harnesses the interference of two laser frequency combs to provide unprecedented capability in spectroscopy applications. In the past decade, the state-of-the-art systems have reached a point where the signal-to-noise ratio per unit acquisition time is fundamentally limited by shot noise from vacuum fluctuations. To address the issue, we propose an entanglement-enhanced dual-comb spectroscopy protocol that leverages quantum resources to significantly improve the signal-to-noise ratio performance. To analyze the performance of real systems, we develop a quantum model of dual-comb spectroscopy that takes practical noises into consideration. Based on this model, we propose quantum combs with side-band entanglement around each comb lines to suppress the shot noise in heterodyne detection. Our results show significant quantum advantages in the uW to mW power range, making this technique particularly attractive for biological and chemical sensing applications. Furthermore, the quantum comb can be engineered using nonlinear optics and promises near-term experimentation.
\end{abstract}

\date{\today}
\maketitle

\section{Introduction}
Dual-comb interferometry, a frequency-comb-based precision measurement technique harnessing the interference of two laser frequency combs of slightly different repetition rates in a static device, has emerged to provide unprecedented  capability in various applications including spectroscopy~\cite{Picque2019, Coddington:16, Fortier2019}, hyperspectral imaging \cite{Martin-Mateos_imagingOptica, Vicentini2021}, and light detection and ranging (LiDAR)~\cite{Coddington2009RapidAP, kippenberg_ranging2011, Lukashchuk2022, Muh_ranging2018, Caldwell2022}. In combination with on-chip frequency comb generators, dual-comb technique has been demonstrated with various platforms, including quantum cascade lasers \cite{Villares2014,yang2016terahertz}, micro-resonator-based soliton combs \cite{Suhscience2016, Yu_Nat_commu_2018}, electro-optic micro-rings~\cite{Shams-Ansari2022} and on-chip semiconductor lasers~\cite{van2020chip}.

In terms of spectroscopy, dual-comb interferometry has unique advantages of (1) rapid effective data acquisitions without mechanical moving parts \cite{Ideguchi2013, Bernhardt2010, Ideguchi:16}; (2) broad spectral coverage over the large span of a comb generator~\cite{Muraviev2018, Chen2018, Ycas2018, AbijithScience_advance_2019}; (3) spectral resolution reaching the comb line spacing of sub-picometers~\cite{Chen2018, CoddingtonPhysRevLett.100.013902}; (4) frequency scale calibrated with the accuracy of an atomic clock \cite{Chen2019PNAS}; (5) feasibility to on-chip integration \cite{Suhscience2016, Yu_Nat_commu_2018}, whose high repetition rates supports rapid measurements. For linear spectroscopy, it acquires thousands of molecular transitions simultaneously \cite{Muraviev2018, Chen2018, Ycas2018}, providing rich spectral information for quantitative concentration analysis on a sample~\cite{Rieker:14, Coburn:18, breath_analysis}.  For nonlinear spectroscopy, the high peak power of the comb sources has been utilized for coherent anti-stokes Raman spectroscopy \cite{Ideguchi2013} and  four-wave-mixing multi-dimensional spectroscopy \cite{Lomsadze_multidimension_science} on bio-chemicals, with the potential to improve the measurement speed by several orders of magnitude.

Similar to other broadband spectroscopic techniques,
the sensitivity in a dual-comb measurement is inversely proportional to the optical bandwidth as the laser power in photo-detection is constrained due to detector nonlinearity or sample damage~\cite{Coddington:16, Newbury:10}. In this regard, the product of signal-to-noise ratio (SNR) per unit acquisition time and the number of resolved spectral elements is taken as the figure of merit for a dual-comb system~\cite{Coddington:16}. 
As long-term coherent averaging  \cite{CoddingtonPhysRevLett.100.013902, Chen2018, Ideguchi2014, Millot2016} improves the sensitivity at the cost of acquisition time~\cite{Muraviev2018, Chen2018, Ycas2018, AbijithScience_advance_2019}, these techniques hinder real-time sensing and fail to improve the figure of merit. \QZ{Indeed, one has to suppress various noise sources to achieve a higher figure of merit, including detector noises, laser relative intensity noise (RIN), shot noise and others. While more device engineering efforts can potentially suppress detector noise and RIN, the fundamental shot noise remains. In this regard, state-of-the-art dual-comb sensing systems have shown examples where shot noise is dominant~\cite{zolot2012direct,baumann2011spectroscopy,kowligy2019shot}.}

To go beyond the shot noise limit, quantum resources such as squeezing and entanglement are necessary. For example, the Laser Interferometer Gravitational-wave Observatory (LIGO)~\cite{abadie2011gravitational,aasi2013enhanced,tse2019quantum} and the Haloscope At Yale Sensitive To Axion CDM (HAYSTAC) dark matter search~\cite{backes2021} inject squeezed light to suppress the shot noise. Photonic radar~\cite{xia2020demonstration} and optomechanical force sensing~\cite{xia2023entanglement} adopted entanglement with the distributed sensing paradigm~\cite{zhang2021distributed}. In terms of spectroscopy, amplitude-squeezing has been demonstrated in nonlinear spectroscopy~\cite{casacio2021quantum} and entangled two-mode squeezed vacuum has been shown to benefit linear absorption spectroscopy~\cite{shi2020}. However, none of these advantages directly applies to dual-comb spectroscopy, as its essential component of heterodyne detection presumably precludes the use of squeezing and entanglement. 

In this work, we develop a quantum description of dual-comb spectroscopy and then propose an entanglement-enhanced scheme that utilizes quantum combs of light to gain sensitivity enhancement in dual-comb spectroscopy. We first provide a complete quantum model for dual-comb spectroscopy, which recovers the SNR results of Ref.~\cite{Newbury:10} in the case of classical source. Furthermore, the quantum model allows us to design a quantum comb composed of pair-wise entanglement around each strong comb line to improve the SNR drastically. The quantum advantage is robust against loss and phase misalignment. We also provide an experimental design to engineer the quantum comb with off-the-shelf components.

\section{Results}
\subsection{Overview of the protocol}
\begin{figure}
    \centering
    \includegraphics[width=\linewidth]{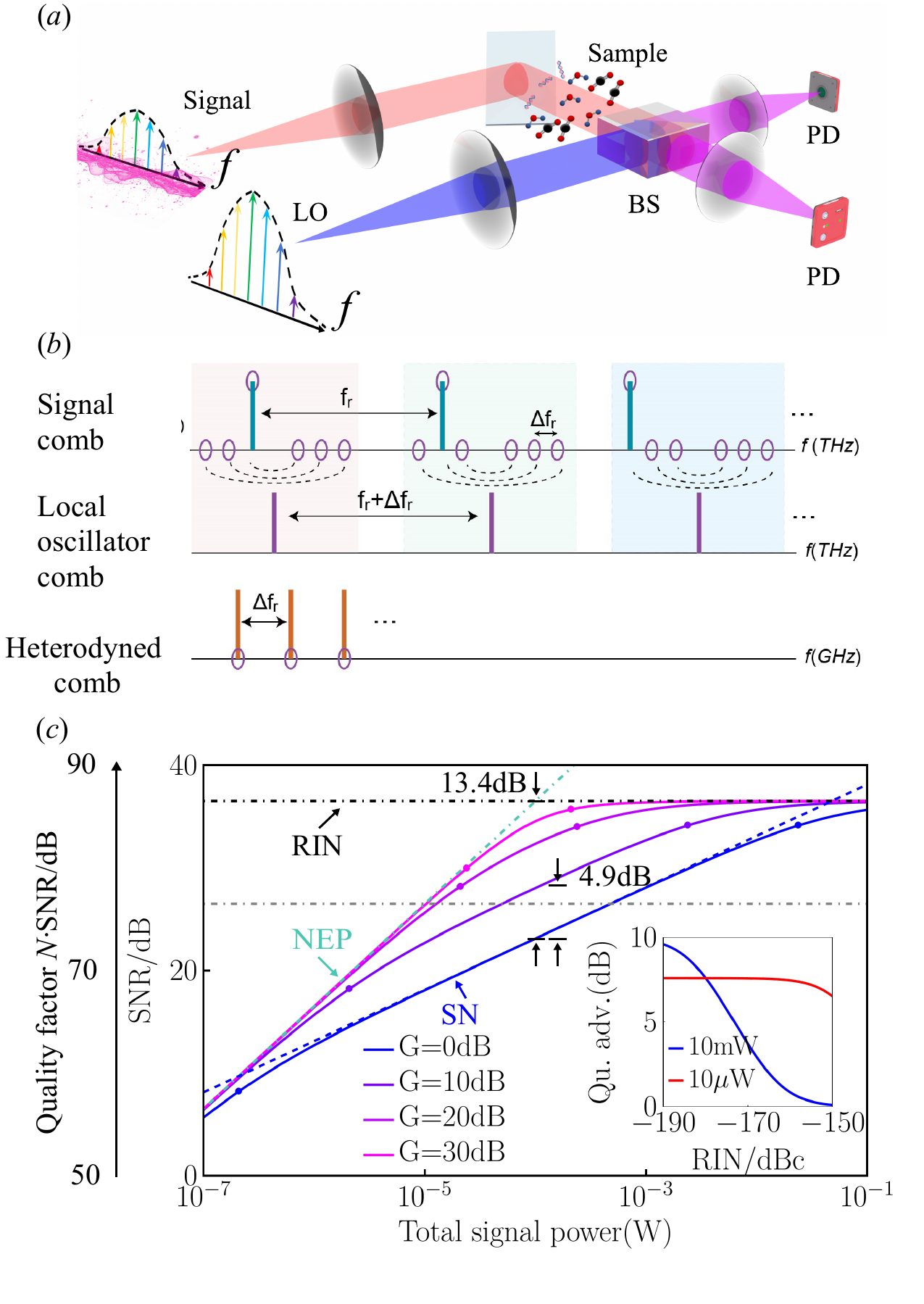}
    \caption{
(a) Conceptual schematic for entanglement-enhanced dual comb spectroscopy. The teeth share intermodal entanglement within the signal comb (red beam). LO: local osillator; BS: beamsplitter; PD: photon detector.
(b) Schematic of the quantum comb. Each pair of signal modes beating with the same LO comb tooth (purple line) for the same intermediate frequency is entangled, indicated by a black dashed line connecting a pair of purple circles.
(c) Practical SNR involving NEP-type and RIN-type noises, plotted versus signal power (analog to Fig.~2 of Ref.~\cite{Newbury:10}), normalized to unit acquisition time $T=1s$. We assume an ideal detector with unity efficiency, and zero loss and noise $\kappa\approx 1, \eta=1$. In (c), both signal and LO are entangled with equal gain $G$, which increases from $0$dB (coherent-state) to $30$dB in steps of $10$dB, plotted in color from blue to magenta. The NEP/RIN-dictated SNR is presented by green-dot-dashed/black-dot-dashed line, along with the shot noise (SN) limit in blue-dashed. $ N=10^5, \lambda=1\mu$m, RIN$=-170$dBc/Hz~\cite{zhou2020mid} (a more accessible RIN$=-150$dBc/Hz enforces an earlier saturation shown by the gray dot-dashed line), NEP$=5\times 10^{-13}$W/Hz$^{1/2}$ (NEP$=4.5\times 10^{-15}$W/Hz$^{1/2}$ is actually achievable, e.g. by Thorlabs FGA01FC-InGaAs Photodiode), $P_{\rm LO}/P_{\rm S}=5$. Inset: Quantum advantage in SNR (in decibel unit) versus various values of RIN for total signal power $P_{\rm S}$=10mW (blue) and $P_{\rm S}$=10$\mu$W (red) at $G$=20dB. The LO-signal power ratio $\gamma\equiv P_{\rm LO}/P_{\rm S}=5$ is fixed---although this figure shows the case of both signal and LO entangled, as we show later, only the signal needs to be entangled under large $\gamma$. 
    }
    \label{fig:overview}
\end{figure}

Dual-comb spectroscopy employs the interference between the signal comb (shown in red) and the local comb (shown in blue), as shown in Fig.~\ref{fig:overview}(a). The protocol's strength stems from the selection of signal and local combs with slightly different frequency spacings, $f_r$ and $f_r+\Delta f_r$, respectively, as illustrated in Fig.~\ref{fig:overview}(b). The signal comb interrogates the sample and undergoes loss and phase shift that can be modeled as a bosonic quantum channel, with frequency-dependent transmissivity $\kappa(f)$ and phase-shift $\alpha(f)$. Meanwhile, the local comb serves as a local oscillator (LO) for the final heterodyne measurement. After mixing the LO and return at a balanced beamsplitter, information about the sample can be extracted from the photocurrent difference $\hat N(t)=\hat c_+^\dagger (t)\hat c_+(t)-\hat c_-^\dagger (t) \hat c_-(t)$, obtained from the photocurrent measurement on both output ports $\hat c_{\pm}(t)$. While we utilize quantum operator language to describe the measurement to prepare our analyses for quantum combs, our analyses also recover the results obtained from semi-classical analyses for classical combs~\cite{Newbury:10}.

In the classical protocol, both combs are classical--- the quantum state of each comb line is in a coherent state, obeying the shot-noise-limited standard quantum limit (SQL).
We propose to engineer quantum combs to further improve the performance of dual-comb spectroscopy. To begin with, we consider the case of signal comb being quantum engineered. To suppress noise below the SQL, squeezing is commonly adopted in quantum sensing protocols. However, in the case of dual-comb acquisitions, squeezing a single mode alone is inadequate, as heterodyne measurement is necessary to read out quadratures across the entire spectrum. To surpass the SQL, entanglement between different frequency modes is required so that joint quadratures are squeezed. Thus, we propose entangling the side-bands of the signal comb around each local comb line, as indicated in Fig.~\ref{fig:overview}(b) by the dashed lines. Such an entangled comb  with squeezing gain $G$ allows both measured quadratures in heterodyne to be squeezed, resulting in fundamental noise a factor of ${1}/{G}$ below the SQL 
Furthermore, as we will detail in later part of the paper, in general, the LO can be engineered to be similarly entangled, which can further improve the performance, especially when the power of the local comb is similar as or lower than that of the signal comb. Note that quantum engineering almost preserves the power of the comb, as the joint-squeezing power is negligible compared to the mean field in dual-comb systems.

To analyze the performance of the proposed quantum dual-comb spectroscopy protocol, we model the signal and all noise involved in the protocol systematically. In this overview, the case of bright LO is considered, where the power $P_{\rm LO}$ is much larger than the signal comb power $P_{\rm S}$, while the full analysis is presented later. The information about the sample is derived from the amplitude decay and phase-shift of the signal comb, $e^{i\alpha}\sqrt{\kappa P_{\rm S}}$, which is subject to contamination from various sources of noise. For low-loss ($\kappa\sim1$) and low-noise scenarios, typical for room-temperature dilute chemical gas sensing and bio-sensing with thin sample slices at signal wavelength $\lesssim 5$um, the fundamental noise stemming from fluctuation properties of the light field $\sim 1/2G$ for the quantum comb. In terms of estimating the transmissivity $\sqrt{\kappa}$, it contributes an inverse-law noise term $O(1/P_{\rm S})$. In practice, the detector noises characterized by noise equivalent power (NEP) and the relative intensity noise (RIN) in laser sources also mix in, which dominate the estimation error at the low and high comb power region respectively, as analyzed for the classical protocol in Ref.~\cite{Newbury:10}. In addition to the inverse-law term due to the fundamental noise, the RIN adds a constant noise term $O(1)$ independent of $P_{\rm S}$, denoted as RIN-type noise, and the NEP adds an inverse-square-law noise term $O(1/P_{\rm S}^2)$ to the estimation, denoted as NEP-type noise. We ignore detector dynamical range noise, as it can be resolved by engineering the detector array~\cite{Newbury:10} and it has similar effects as RIN-type noise. 

With all noises into consideration, we consider the realistic performance of the quantum dual-comb spectroscopy system. 
\QZ{For generality, our calculation covers a wide comb power range from 10$^{-7}$ W to 10$^{-1}$ W, the same as that in Ref.~\cite{Newbury:10}. Indeed, in a typical dual-comb application (power ratio $\gamma$=1), the total comb power should be constrained below 50 $\mu$ W in the near-infrared range ~\cite{Chen2018} and below 30 $\mu$W in the mid-infrared region~\cite{Chen2019PNAS,baumann2011spectroscopy} to avoid detector nonlinearity. This is, most of the time, due to the  requirement of a high dynamic range to sample the strong center burst in the dual-comb interferogram. A promising route of engineering a high dynamic range detector has been able to extend to operation power up to 2 mW without clear spectral artifacts observed, as demonstrated in Ref.~\cite{Roy:12}. }


We plot the signal-to-noise ratio (SNR) per second versus the total signal power under practical experimental settings in Fig.~\ref{fig:overview}(c). In this uW-mW region, the performance of state-of-the-art classical systems (blue solid) is limited the shot noise (SN) limit (blue dashed). While at low/high signal power limit, the SNR converges to the NEP/RIN-dictated limit (green dot-dashed/black dot-dashed). With the quantum comb, we see that a practical entangled source of $10$dB gain ($G=10$, purple solid) yields a quantum advantage up to 4.9dB over the coherent-state source (blue solid). As the gain increases (from blue to magenta), the quantum advantage improves further until it saturates subject to the limits dictated by NEP-type noise alone (greed dot-dashed) and the RIN-type noise alone (black dot-dashed). In the scenario of Fig.~\ref{fig:overview}(c), we observe that the ultimate limit of quantum advantage can go up to $13.4$dB at $P_{\rm S}\approx 0.1$mW, which is of great interest to bio-sensing applications. Additionally, we provide predictions at the power levels when such saturation happens (dots on the blue-magenta curves), as we detail in Appendix~\ref{app:SNR_full}. Our analyses show that for a state-of-art dual-comb system to enjoy quantum advantage, RIN-type of noise is often the major constraint: in the inset of Fig.~\ref{fig:overview}, we show that for 10mW signal power, to enjoy a significant quantum advantage it requires RIN-type noise to be smaller than $\sim -170$dBc/Hz, challenging but still possible~\cite{morton2018high}; For lower power of 10uW, the requirement is less stringent, RIN$=-150$dBc/Hz is readily achievable~\cite{Newbury:10,zhou2020mid}.

\subsection{General amplitude and phase detection}
In a general spectroscopy sensing process, one is interested in the input-output relation of light for a range of frequencies $f$. The pattern of the output light reveals information about the composition of the sample under study. The mathematical model for the input-output relation involves a thermal-loss phase-shift channel, which has frequency-dependent transmissivity $\kappa(f)$ and phase-shift $\alpha(f)$. Given an input light mode described by the annihilation operator $\hat{a}_S$~\footnote{Annihilation operators satisfy canonical commutation relation $[\hat{a}_S^\dagger,\hat{a}_S]=1$.}, the output field annihilation operator is given by the linear relation
\be 
\hat a_R=\sqrt{\kappa}e^{i\alpha}\hat a_S+\sqrt{1-\kappa}\hat a_E.
\label{a_input_output} 
\ee
The channel attenuates the mean of input signal mode $\hat a_S$ by $\sqrt{\kappa}$, shift the phase by $\alpha$, and mixes in the environment mode $\hat a_E$ with mean thermal photon number given by the Bose-Einstein distribution
$
\calE(f)=1/[\exp\!\left(h f /k_B T_B\right)-1],
$
with $h$ being the reduced Planck constant, $k_B$ the Boltzmann constant and $T_B$ being the sample environment temperature. Although the thermal noise $\calE(f)\ll 1$ is negligible at the frequency of interest, we will keep it in our analyses to tackle the general case.

Although our results work for the simultaneous estimation of phase-shift and transmissivity, as enabled by dual-comb technique, we consider two special scenarios to simplify the SNR analyses. In the first scenario, we are concerned with only the transmissivity $\kappa(f)$, while the phase-shift is negligible due to phase cancellation via sending both combs to the sample. In the second scenario, the absorption is almost zero ($\kappa(f)\simeq 1$), while the phase-shift $\alpha(f)$ (despite also being small) provides the major information. For example, imaging the subtle changes in phase contrast on the order of 0.1 milli-radians allows the study of neural activities at the single neuron level~\cite{PNAS_neuron_activity, Park2018}. Overall, the absorption or phase-shift can be very weak due to the low concentration of sample, as is the case in atmospheric sensing and human breath analysis~\cite{breath_analysis} at parts per billion and in radiocarbon detection at few parts per quadrillion~\cite{Galli:16}.

In our analyses, we will ignore phase noise, since various dual-comb noise suppression techniques have been developed, based on comb-source engineering \cite{Yang2017, Avik_dual_comb, Fritsch2022}, digital phase correction \cite{Ycas2018, Ideguchi2014, Roy:12}, and active stabilization \cite{Chen2018}, with some of them achieving coherent times up to hours \cite{Chen2018, Ycas2018, Roy:12}.

\subsection{Quantum model of dual comb spectroscopy}
Now we formulate the quantum theory for dual comb spectroscopy of $N$ frequency components. In the rotating frame of the carrier frequency $\nu_0$ ($\nu_0\gg f_r$), the signal comb is represented by the field operator
\be 
\hat A(t)= \frac{1}{\sqrt T}\left[
\hat a(t) +\sum_{n=1}^N
A_n e^{i 2\pi n(f_r+\Delta f_r)t} \right],
\label{eq:fullinput_A}
\ee
while the local oscillator (LO) comb is represented by
\be
\hat B(t)=\frac{1}{\sqrt T}
\left[\hat b (t)+ \sum_{n=1}^N B_n e^{i 2\pi n f_r t}\right].
\label{eq:fullinput_B}
\ee
Here $T$ is the acquisition time, and $f_r>1/T$ to avoid aliasing. The sum in each comb consists of the strong mean fields of a frequency comb source at discrete frequencies. The light power is mainly contributed by these mean fields. Specifically, the power is $P_{\rm S} =h\nu_0 \sum_{n=1}^N |A_n|^2/T$ for the signal, and $P_{\rm LO} =h\nu_0 \sum_{n=1}^N |B_n|^2/T$ for the LO, while the additional power due to squeezing is negligible in this paper. The quantum-operator term in each comb describes the noise
\begin{align}
\hat z(t)=\sum_{n=1}^N \sum_{{\delta}=-N}^{N}\hat z_{n,{\delta}} e^{i 2\pi [n f_r + {\delta} \Delta f_r] t}
\end{align}
where $\hat z\in \{\hat a,\hat b\}$, and we have quantized the frequency modes of field into the field annihilation operators, satisfying the commutation relation $[\hat a_{n,{\delta}},\hat a_{n,{\delta}}^\dagger]=[\hat b_{n,{\delta}},\hat b_{n,{\delta}}^\dagger]=1$ and all the other commutators are zero. \QZ{The double subscripts $n\in[1,N]$ and $\delta\in[-N,N]$ determines the frequency of the mode $n f_r + {\delta} \Delta f_r$---$n$ denotes which comb line the mode is around, while $\delta$ further specifies which mode around the comb line specificed by $n$. As $\Delta f_r \ll f_r$, the sideband modes around different comb lines will not overlap.} 
Here we have included all frequency modes relevant to the heterodyne measurement. In a classical strategy, the noise property of all modes is vacuum-limited, but in this work, we propose to engineer the noise property via squeezing and entanglement.

\begin{figure}[t]
    \centering
    \includegraphics[width=0.9\linewidth]{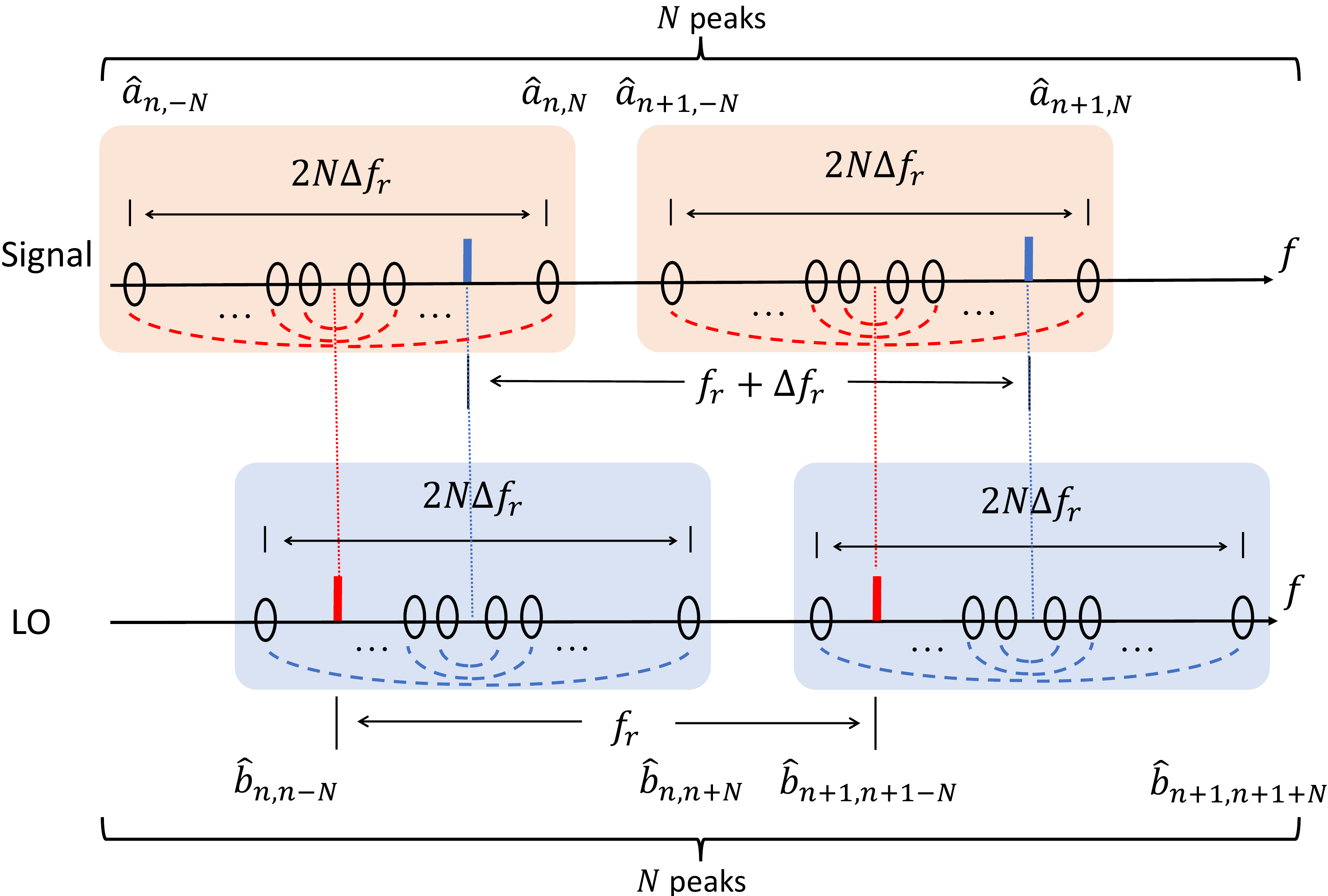}
    \caption{The frequency arrangement of the comb modes. \QZ{A red peak $B_n$ in LO beats with the side band modes in the signal (connected by red dashed lines), whose quadrature fluctuation contributes to the overall noise. Therefore, we entangle the side bands (connected by red dashed lines) of the signal to improve the SNR.
    Similarly, a blue peak $A_n$ in signal beats with the sideband mode pairs in LO connected by blue dashed lines, which are to be entangled to improve the SNR.} Around each peak, there are $N$ such pairs. Here we explicitly label the pairs at the edges. }
    \label{fig:comb_freq_position}
\end{figure}

Field propagation through the sample can be formulated by a bosonic quantum channel, as shown in Eq.~\eqref{a_input_output}. We are interested in the transmissivity $\kappa(f)$ and phase $\alpha(f)$ induced by the sample. Note that the non-ideal LO storage also induces a channel of transmissivity $\eta(f)$ and phase-shift $\beta(f)$. We assume that the transmissivity and phase-shift spectra are smooth enough such that their values at sidebands of each comb line are identical. For example, $\kappa(nf_r+{\delta}\Delta f_r)= \kappa_n$ for all sideband frequencies $-N \! \le \! \delta \! \le \! N$. 
We define $\alpha_n$ and $\eta_n, \beta_n$ similarly. Our formalism can be easily generalized to rapidly-varying spectra, while the formula will turn much lengthier. After travelling through the sample, channel output fields $\hat A^\prime (t)$ for the sample return and $\hat B^\prime (t)$ for the LO can be decomposed in the same form as Eqs.~\eqref{eq:fullinput_A} and~\eqref{eq:fullinput_B}. According to Eq.~\eqref{a_input_output}, the input-output relation yields 
$A_n\to \sqrt{\kappa_n}A_ne^{i\alpha_{n}}$ and $B_n\to \sqrt{\eta_n}B_ne^{i\beta_{n}}$ for mean fields, and
\bal 
\hat a_{n,{\delta}}^\prime&= \sqrt{\kappa_{n}}e^{i\alpha_{n}} \hat a_{n,{\delta}} +\sqrt{1-\kappa_{n}} \hat e_{n,{\delta}}  \,, \\
\hat b_{n,{\delta}}^\prime&= \sqrt{\eta_{n}}e^{i\beta_{n}} \hat b_{n,{\delta}} +\sqrt{1-\eta_{n}} \hat f_{n,{\delta}}  \,,
\label{eq:IOrelation}
\eal 
for noise modes,
where $\hat e_{n,{\delta}}$'s and $\hat f_{n,{\delta}}$'s are environmental noise modes of thermal photon number $\calE_{n}\equiv \calE(nf_r)$.

At the receiver, the two combs are combined by a balanced beamsplitter, yielding two output combs, 
$
\hat c_\pm(t)=[ \hat A^\prime(t) \pm \hat B^\prime(t)]/\sqrt{2}
$.
Then the photon counts of the two output combs are measured, and subtracted from each other $
\hat N(t)=\hat c_+^\dagger (t)\hat c_+(t)-\hat c_-^\dagger (t) \hat c_-(t)
=\hat A^{\prime\dagger}(t)\hat B^\prime(t)+\hat B^{\prime\dagger}(t)\hat A^\prime(t)$. Taking into account that $\Delta f_r\ll f_r$,  we can filter out the direct current (DC) term and the fast-oscillating terms at $|f|\gg \Delta f_r$. The resulting alternative current (AC) $ N_{\rm AC}(t)$ is a random variable with mean
\begin{align}
 \overline N_{\rm AC}(t)=\frac{1}{T}\left[\sum_{n=1}^N\sqrt{\kappa_n \eta_n} e^{i(\alpha_n-\beta_n)} A_n B_n e^{i 2\pi n\Delta f_r t}+c.c.\right]\,,
 \label{NAC_t}
\end{align}
where $c.c.$ represents the complex conjugate.
One can perform a finite-time-$T$ Fourier transform to obtain the spectrum
\be 
\overline N_{\rm AC}(m\Delta f_r)=
\sqrt{\kappa_m\eta_m} A_{m} B_{m}  e^{i(\alpha_m-\beta_m)}, 1\le m \le N,
\label{eq:Nac_mean}
\ee 
from which we can extract the information about the transmissivities and phase-shifts across the entire $N$-line spectrum.

To evaluate the fluctuation of the readout, now we consider the contribution to $N_{\rm AC} (m\Delta f_r)$ from noise modes $\hat a, \hat b$. As the amplitudes $A_n,B_n\gg 1$, the noise in $ N_{\rm AC}(m\Delta f_r)$ is
\begin{align}
&\hat \Sigma_{\rm AC}(m\Delta f_r)\simeq  \sum_{n=1}^N \left[\sqrt{\eta_n \kappa_n}B_n \hat X_{n,m} + \sqrt{\eta_n \kappa_n}A_n \hat Q_{n,m}\right.
\nonumber
\\
&\left.
+
\sqrt{\eta_n (1-\kappa_n)}B_n \hat X^{(e)}_{n,m}+ \sqrt{(1-\eta_n) \kappa_n}A_n \hat Q^{(f)}_{n,m} \right].
\label{eq:Nac}
\end{align}
For the full derivation of Eqs.~\eqref{eq:Nac_mean}\eqref{eq:Nac}, see Appendix~\ref{app:derivation}. Here we have adopted the nomenclature widely used in quantum optics~\cite{bowen2015quantum,malnou2019squeezed} that defines the joint quadrature operators
\bal 
\hat X_{n,m}&\equiv \hat a_{n,m}e^{i(\alpha_n-\beta_n)}+\hat a_{n,-m}^\dagger e^{-i(\alpha_n-\beta_n)}\,, 
\\
\hat Q_{n,m}&\equiv \hat b_{n,n+m}e^{-i(\alpha_n-\beta_n)}+\hat b_{n,n-m}^\dagger e^{i(\alpha_n-\beta_n)}\,, 
\label{eq:quadrature_def}
\eal 
for the signal ($\hat a_{n,m},\hat a_{n,-m}$ beat with the strong mean field $B_n$ at frequency $nf_r$) and for the LO ($\hat b_{n,n+m},\hat b_{n,n-m}$ beat with $A_n$ at frequency $n(f_r+\Delta f_r)$) respectively. 
Simiarly, we define the quadratures $\hat X^{(e)}, \hat Q^{(f)}$ for the environment modes $\hat e$ and $\hat f$ in Eq.~\eqref{eq:IOrelation}. Note that these quadratures, along with $\hat\Sigma_{\rm AC}$, are usually not Hermitian (real-valued) observables, thus their variances are defined as $\text{var} \hat X\equiv \expval{\hat X^\dagger \hat X}$ for any non-Hermitian complex operator $\hat X$.

Dual-comb spectroscopy aims to estimate the transmissivity $\kappa_n$, phase-shift $\alpha_n$ or both simultaneously, for all $1\le n \le N$ frequencies of the sample from the photo-current of Eq.~\eqref{eq:Nac_mean}. We define the amplitude SNR at each comb line as 
\be 
\text{SNR}=|\overline N_{\rm AC}(m\Delta f_r)|/\sqrt{{\rm var}\left[N_{\rm AC}(m\Delta f_r)\right]}\,.
\label{eq:SNR_bothquad}
\ee 
The noise, which is defined in Eq.~\eqref{eq:Nac}, collects the beating modes near all $N$ comb lines. As shown in Appendix~\ref{app:justification_SNR}, it is a good indicator for the minimum mean square error of either the transmissivity or the phase-shift estimation task. Furthermore, a neat figure of merit is the overall quality factor $N\cdot \text{SNR}$, which eliminates the dependence on total line number $N$. 

To evaluate the SNR, we make use of the independence between modes around different comb lines and evaluate the variance from Eq.~\eqref{eq:Nac}, 
\begin{align}
&{\rm var}\left[\hat \Sigma_{\rm AC}(m\Delta f_r)\right]=\sum_{n=1}^N \Big[\calN_n
\nonumber
\\
&\quad\quad\left.
+\eta_n \kappa_n \left(B_n^2{\rm var}\hat{X}_{n,m}+A_n^2{\rm var}\hat{Q}_{n,m} \right)\right],
\label{eq:var_NAC_full}
\end{align}  
where the thermal noise $\calN_n=\eta_n B_n^2(1-\kappa_n)(2\calE_n+1)+\kappa_n A_n^2 (1-\eta_n)(2\calE_n+1)$ is determined by the sample, the LO storage and the environment temperature; the complex quadrature noises ${\rm var}\hat{X}_{n,m}$ and ${\rm var}\hat{Q}_{n,m}$ are determined by the quantum state of the signal comb source and local comb source.

\subsection{SNR with entangled quantum comb}
From the definition of quadratures in Eq.~\eqref{eq:quadrature_def}, we see that the noise ${\rm var}\hat{X}_{n,m}$ can be suppressed by entangling the modes $\hat a_{n,\pm m}$ in a two-mode squeezed vacuum state (see Appendix~\ref{app:TMSV}). By such means, the joint quadrature $\hat{X}_{n,m}$ is squeezed with suppressed variance
\be 
{\rm var}\hat{X}_{n,m}=\frac{1}{ 2G}\left[-\left(G^2-1\right) \cos \left(2 \alpha_n-2 \beta_n\right)+(G^2+1)\right],
\label{eq:varX_TMSV}
\ee 
where squeezing gain $G\ge 1$. When phases are perfectly matched as $\alpha_n-\beta_n=0$, ${\rm var}\hat{X}_{n,m}$ is minimized to $1/G $.
Similarly, we can squeeze the joint quadrature $\hat{Q}_{n,m}$ of the local comb by gain $G_{\rm LO}$.
When $G=G_{\rm LO}=1$, the variance reduces to the classical dual-comb spectroscopy. In this case the variance of complex operator ${\rm var}\hat{X}_{n,m}$ is twice of the SQL $1/2$, because it is defined as a sum of variances of its real and imaginary parts.

To model the full SNR of the dual-comb spectroscopy system, we involve device and source imperfections. For simplicity, we assume the $N$ comb lines are generated symmetric ($A_n=A_{n'}$ and $B_n=B_{n'}$). In Appendix~\ref{app:SNR_full}, we derive the full formula of the SNR at intermediate frequency $m\Delta f_r$
\begin{align}
&\text{SNR}^{-2}=
& \frac{ N^2}{T} \left( a_\text{NEP}\frac{1}{P_{\rm S}^2}+\frac{a_\text{quad}  }{P_{\rm S}}+a_\text{RIN} \right),
\label{SNR_overall}
\end{align} 
where $T$ is the acquisition time, $P_{\rm S}$ is the total signal power, the NEP-type noise coefficient $a_\text{NEP}\equiv \text{NEP}^2/\eta_m\kappa_m\gamma$, the RIN-type noise coefficient $a_\text{RIN}\equiv \text{RIN}/2$, and the quadrature noise coefficient
\be 
a_\text{quad}\equiv \frac{h\nu_0 }{N} \sum_{n=1}^N \left[  \left({\rm var}\hat{X}_{n,m}+\frac{1}{\gamma} {\rm var}\hat{Q}_{n,m} \right)+\frac{\calN_n}{\kappa_n\eta_n }\right].
\label{eq:a_quad}
\ee 
Here $\gamma\equiv P_{\rm LO}/P_{\rm S}$ is the LO-to-signal power ratio, $h\nu_0$ is the energy per photon. Note the quadrature noises can be suppressed by the entanglement (joint quadrature squeezing) in Eq.~\eqref{eq:varX_TMSV}. The proposed SNR quantifies the performance of both the transmissivity estimation and phase estimation scenarios.

The SNR of Eq.~\eqref{SNR_overall} versus total signal power $P_{\rm S}$ has been evaluated in Fig.~\ref{fig:overview}(c) for the case of $G=G_{\rm LO}>1$, which highlights the quantum advantages from entanglement. We have taken the case where the phase mismatch $\alpha_n-\beta_n\ll 1/G$ are all small. This is the case when one estimates the transmissivity with good phase locking, or estimates small phase-shift caused by weak samples---phase-shift as small as 0.1 milli-radians allows the study of neural activities at the single neuron level~\cite{PNAS_neuron_activity, Park2018}. \QZ{When there is phase noise, then $\alpha_n-\beta_n$ cannot be set to zero. Suppose there is mismatch $\epsilon \ll1$, then the leading order of the variance in Eq.~\eqref{eq:varX_TMSV} ${\rm var}\hat{X}_{n,m}=1/G+2\epsilon^2G$ will mix in the anti-squeezing part. Therefore, phase-locking is crucial for the proposed EA dual-comb spectroscopy, similar to other squeezing enabled protocols. }

Our formula can be regarded as a quantum version of Eq.~(2) of Ref.~\cite{Newbury:10}.
We note that our quantum model yields a SNR-$\gamma$ relation different from the semiclassical model in Ref.~\cite{Newbury:10}. Specifically, for a fixed signal comb power $P_{\rm S}$, we find that the optimum is at $\gamma\to \infty$ in our quantum model, while the optimum is finite in the semiclassical model. 
This is because when $\gamma$ is large, the RIN-type noise increases proportional to $\gamma$ in the semiclassical model, while the RIN-type noise remains constant in our quantum model. Our result on the RIN-type noise agrees with Ref.~\cite{wissel2022relative}.

\subsection{Strategies of applying quantum combs}
In the above analyses, we have allowed both the signal comb and local comb to be quantum engineered. In general, having quantum entangled combs in both arms might be not necessary and also experimentally challenging. Here we address different scenarios of applying the quantum combs.

From Eq.~\eqref{eq:var_NAC_full}, we see that the noises from signal and LO are indeed amplified by the mean field of the other, i.e. the signal noise is amplified by the LO mean and vice versa. Hence, squeezing merely the signal is sufficient to yield significant advantage when $P_{\rm LO}$ dominates, while squeezing merely the LO is sufficient when $P_{\rm S}$ dominates.

\begin{figure}
    
    \centering
    \includegraphics[width=\linewidth]{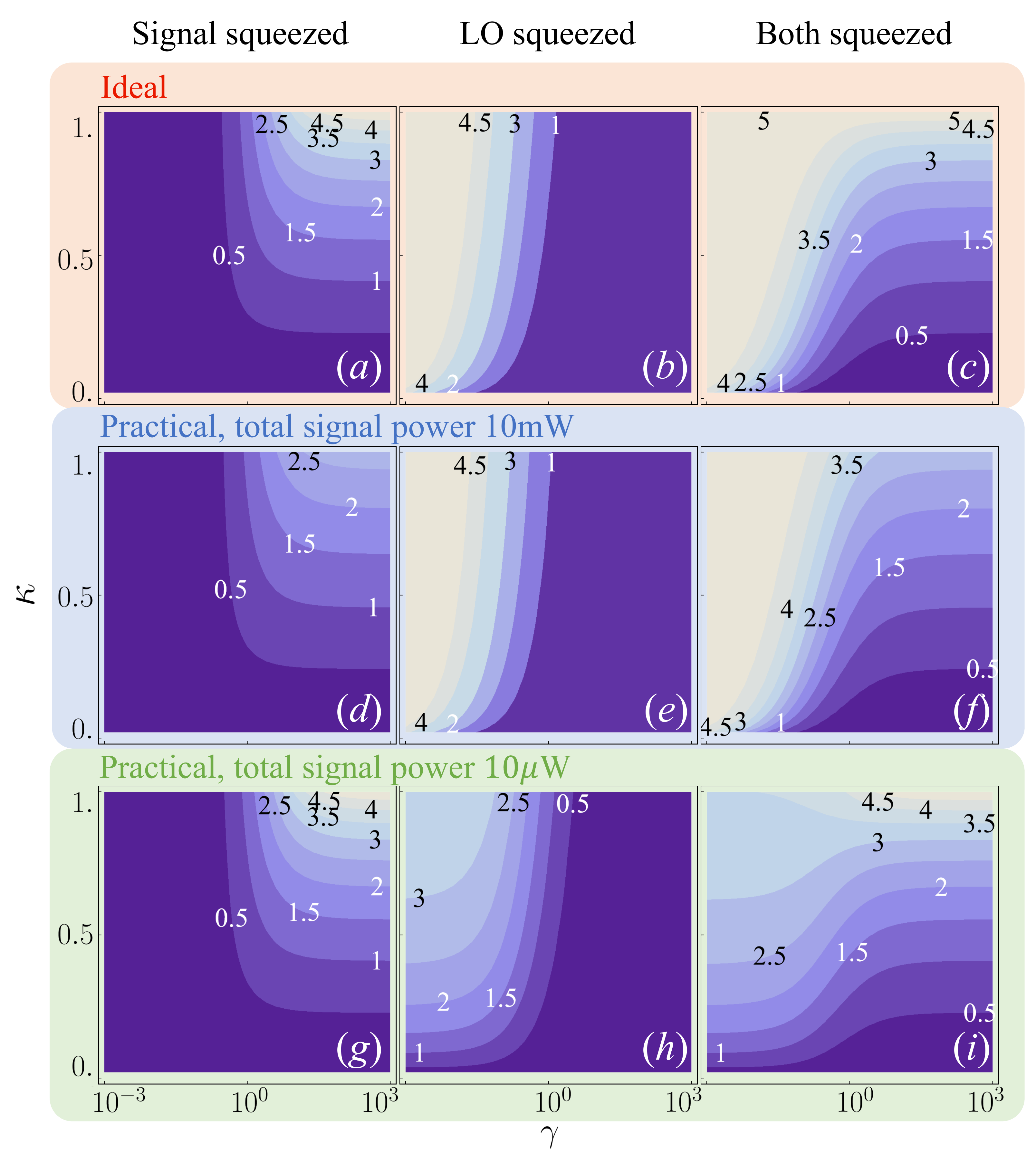}

    \caption{The quantum advantage (in decibel unit) in amplitude SNR over the coherent-state source at the squeezing gain of 10 dB, versus various LO-signal power ratio $\gamma\equiv P_{\rm LO}/P_{\rm S}$ and sample transmissivity $\kappa$ (assumed uniform). Rows: without practical noise; with both NEP-type and RIN-type noises, total signal power $P_{\rm S}=10$mW; with both NEP-type and RIN-type noises, $P_{\rm S}=10\mu$W. Columns:  signal squeezed only;  local comb (LO) squeezed only; both signal and LO squeezed.  We assume ideal detector with unity efficiency and ideal LO link $\eta_m=1$. $N=10^5, T=1s, \lambda=1\mu$m, NEP$=5\times 10^{-13}$W/Hz$^{1/2}$, RIN$=-170$dBc/Hz. 
    \label{fig:contour_LOpower_kappa}
    }
    
\end{figure}

Now we evaluate the quantum advantages of various entanglement (or joint squeezing) strategies over the classical coherent-state source in terms of amplitude SNR. In Fig.~\ref{fig:contour_LOpower_kappa}, row 1  shows the ideal advantage without practical noises (note that the ideal advantage depends on the relative power ratio $\gamma=P_{\rm LO}/P_{\rm S}$ between the signal and LO only, not the absolute magnitudes of power). Row 2 and row 3, operating under signal power $P_{\rm S}=10$mW and $P_{\rm S}=10\mu$W respectively, show the practical advantage with NEP-type and RIN-type noise involved. Along the horizontal axis, all contours are centered at the signal power $P_{\rm LO}=P_{\rm S}$, thus at the left half the signal dominates while at the right half the LO dominates.

In row 1, we verify our results of fundamental limits for the ideal advantages. In Fig.~\ref{fig:contour_LOpower_kappa}(a), only the signal is squeezed, we see that the advantage peaks at the LO-dictated region (right half); in Fig.~\ref{fig:contour_LOpower_kappa}(b), only the LO is squeezed, the advantage now peaks at the signal-dictated region (left half); finally in Fig.~\ref{fig:contour_LOpower_kappa}(c), both the signal and LO are squeezed, here we enjoy both the advantageous areas in the two squeezing strategies above. It is noteworthy that when LO is squeezed, the quantum advantage survives even when $\kappa\to 0$ as shown at the bottom-left corners of subplots (b)(c), which is useful when the sample is lossy and LO power is limited.

In row 2 and row 3, we consider the effect of practical noises. In Fig.~\ref{fig:contour_LOpower_kappa}(d)-(f), the signal power $P_{\rm S}=10$mW is relatively large. In this scenario, the advantage is mainly constrained by the RIN-type noise which is more significant for large $P_{\rm LO}$. We find that the advantages at the LO-dictated region (right half) of all subplots (d)(e)(f) are significantly undermined, which is especially noticeable in subplot (d). On the other hand, in Fig.~\ref{fig:contour_LOpower_kappa}(g)-(i), the signal power $P_{\rm S}=10\mu$W is extremely small. In this scenario, the advantageous region is mainly affected by the NEP-type noise which is more significant for small $P_{\rm LO}$. As expected, the advantages at the left half region of subplots (g)(h)(i) are undermined, which is especially noticeable in subplot (e). In subplot (f) or (i) where both the signal and LO are squeezed, the patterns in the two squeezing strategies shown in subplot (d)(e) or (g)(h) occur simultaneously.

\subsection{Performance under total power constraints}
\begin{figure}
    \centering
 \includegraphics[width=.9\linewidth]{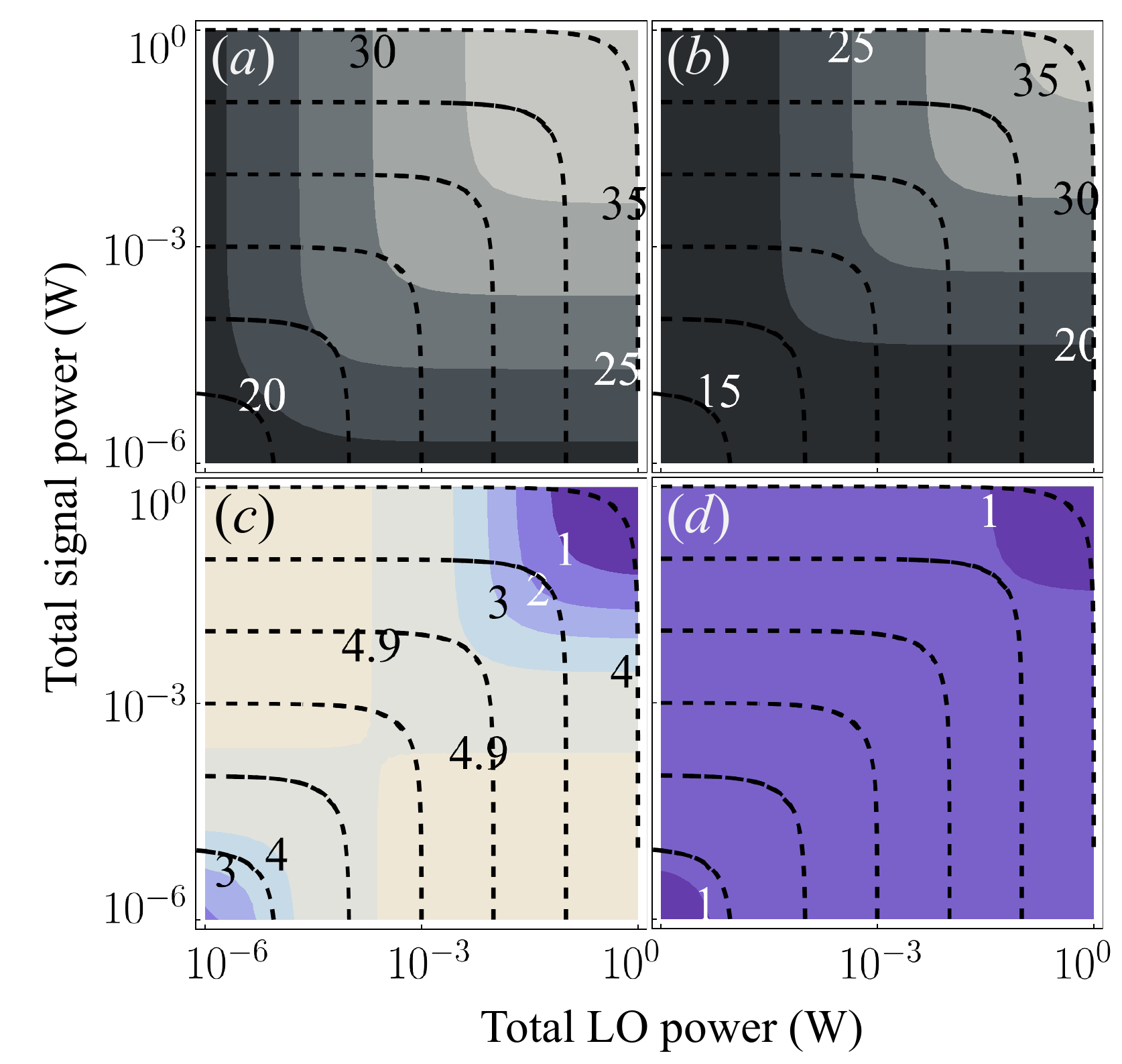}
    \caption{The dependence of (a)(b) the amplitude SNR using the squeezed source and (c)(d) its quantum advantage over the coherent-state source, both in decibel unit, on various power constraints $P_{\rm LO}, P_{\rm S}$. Uniform transmissivity is assumed. Both signal and LO are squeezed at the squeezing gain of 10dB. (a)(c) $\kappa=1$; (b)(d) $\kappa=0.5$. The black dashed lines are contours of total power $P_{\rm LO}+P_{\rm S}$. We assume ideal detector with unity efficiency, while the LO is sent along with the signal so that $\eta_m=\kappa$. $N=10^5, T=1s, \lambda=1\mu$m, NEP$=5\times 10^{-13}$W/Hz$^{1/2}$, RIN$=-170$dBc/Hz. 
    \label{fig:contour_LOpower_Pcpower}
    }
    
\end{figure}
To begin with, we consider the power dependence of the SNR on $P_{\rm LO}$, $P_{\rm S}$. Here we explore the scenario where the LO is sent along with the signal, thus $\eta_m=\kappa$ and the total power exposure $P_{\rm S}+P_{\rm LO}$ is to be constrained. \QZ{Such constraint also naturally appear when detector nonlinearity and saturation is taken into account.}
Fig.~\ref{fig:contour_LOpower_Pcpower}(a)(b) shows that the total power $P_{\rm S}+P_{\rm LO}$ contour  and the SNR contour are tangent at $P_{\rm LO}=P_{\rm S}$.
This explains that in some applications one tends to use comparable LO and signal \QZ{(i.e. the power ratio $\gamma=1$)} rather than very strong LO to save the total power consumption. 

Now consider the quantum advantage. Note that only the fundamental noise is suppressed by the quantum engineering, we expect to maximize these ratios $\sigma^2_{\rm quad}/\sigma^2_{\rm NEP}$ and $\sigma^2_{\rm quad}/\sigma^2_{\rm RIN}$ to see a significant quantum advantage.
In Appendix~\ref{app:SNR_full}, we show that NEP-type noise $\sigma^2_{\rm NEP}\sim 1/(P_{\rm LO} P_{\rm S})$, fundamental noise $\sigma^2_{\rm quad}\sim {(P_{\rm S}+P_{\rm LO})}/{P_{\rm S}P_{\rm LO}}$, RIN-type noise $\sigma^2_{\rm RIN}\sim 1$. The ratio $\sigma^2_{\rm quad}/\sigma^2_{\rm NEP}$ is proportional to the total power $P_{\rm S}+P_{\rm LO}$, while $\sigma^2_{\rm quad}/\sigma^2_{\rm RIN}$ is maximized at $P_{\rm S}\to 0$ or $P_{\rm LO}\to 0$. 

For the NEP-dictated scenario, i.e. the total power is small, the quantum advantage grows with the absolute SNR as total power increases, while it does not depend on $\gamma=P_{\rm LO}/P_{\rm S}$. In this case one can let $\gamma\to 0$ or $\infty$ to make signal or LO dominate so that squeezing on the other is no longer needed, as discussed previously. For the RIN-dictated scenario, i.e. the total power is large, we note that the quantum advantage decreases with the total power and it is  minimized at $P_{\rm S}=P_{LO}$ given a fixed total power, which is opposite to the absolute SNR case. This is not a preferred scenario for quantum advantage. Fig.~\ref{fig:contour_LOpower_Pcpower}(c)(d) verifies that the total power contour and the quantum advantage contour almost overlap in the region of small total power; they are again tangent at $P_{\rm S}=P_{\rm LO}$ in the region of large total power, while the gradient direction of the quantum advantage contour is reversed. 
Comparing subplot (c) of lossless sample $\kappa=1$ and subplot (d) of lossy sample $\kappa=0.5$, we see that the quantum advantage degrades significantly when the sample is lossy now that both signal and LO suffer such loss, while a 1dB advantage still survives.

\subsection{Performance limits in biological applications}
An important application of the proposed entanglement-enhanced dual-comb spectroscopy system is in sensing fragile bio-tissues. In bio-sensing, the power of signal light shining on the sample is typically between $10-100$ uW to avoid kill, bleach or perturb the specimen under analysis~\cite{bass1995handbook,pawley2002confocal,pawley2006handbook}. For example, Ref.~\cite{casacio2021quantum} showed an extreme case of power at $\sim10$mW which causes severe sample damage. In another extreme case retina sensing, the safety standard permits much lower power~\cite{delori2007maximum}. The maximum permissible radiant power is a function of the exposure duration, wavelength and visual angles. In long exposure time limit, the maximum power can be as low as $\lesssim 1$uW (see Fig. 2 of Ref.~\cite{delori2007maximum}). In other scenarios, the power can be higher, e.g. $\sim10-200$uW is adopted in some studies~\cite{yi2015human,schwarz2016safety}. To gain better SNR, therefore one cannot simply increase the power, but rather consider suppressing the quantum-limited noise such as in our proposal.
\begin{figure}[t]
    \centering
    \includegraphics[width=\linewidth]{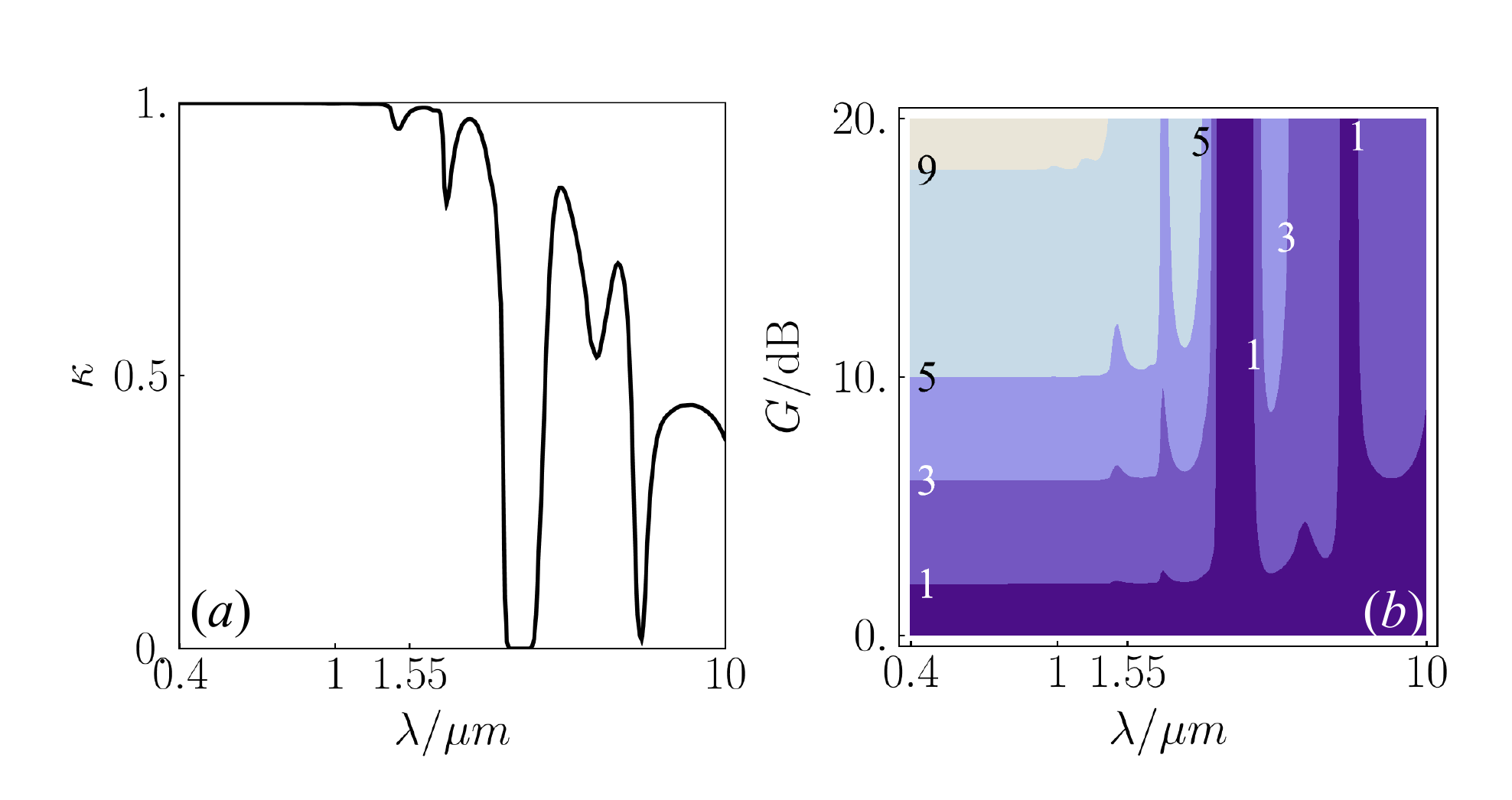}
    \caption{(a) The absorption spectrum of pure water $\kappa(\lambda)$ at room temperature 295K ~\cite{pope1997absorption,kou1993refractive,hale1973optical}, evaluated for path length $L=15\mu$m. (b) The fundamental limit of quantum advantage in amplitude SNR (in decibel unit) enforced by water absorption. The spectrum data subplot (a) is unknown to the observer. An additive thermal Gaussian noise at room temperature is mixed in. $P_{\rm LO}/P_{\rm S}=5$.}
    \label{fig:water}
\end{figure}

In bio-sensing, a major limitation of the applicable frequency region comes from water absorption. The transmissivity spectrum of water absorption can be derived from the Lambert absorption coefficient spectrum $\alpha(f)$ ~\cite{pope1997absorption,kou1993refractive,hale1973optical} via $\kappa(f)=\exp{-\alpha(f) L}$, where $L$ is the sample depth. For the optical domain of wavelength $\lambda<1\mu$m, the absorption is weak: $\alpha\lesssim 10^{-4}/\mu$m. We take a typical sample depth of $L=15$um and evaluate the transmissivity in Fig.~\ref{fig:water}(a) assuming the sample absorption is majorly dominated by water, with absorption coefficients taken from Refs.~\cite{pope1997absorption,kou1993refractive,hale1973optical}. We see that the absorption is substantial starting around $\lambda\sim 2$um. In wavelength below $2$um, the thermal noise described by the Bose-Einstein distribution
$
\calE(f)\ll1
$
is negligible at room temperature. In this $\le 5$um frequency of interest, $\calE\le10^{-4}$ and is negligible at room temperature of $300$K. Even at $10$um, $\calE\sim 0.008$ is still small. With the absorption and noise in hand, we can evaluate the quantum advantage in absence of any NEP-type or RIN-type noise from Eq.~\eqref{eq:a_quad}. To simplify the evaluation, we assume uniform absorption across all comb lines to get a sense of the quantum advantage. In general, the quantum advantage will be an average across a frequency region analysed here. In Fig.~\ref{fig:water}(b), we find the advantage indeed appreciable below about $2$um and increases with the gain $G$, while above $2$um water absorption starts to limit the possible advantage. Note that the quantum advantage monotonically increases with $G$, we expect the contour lines rise to higher $G$ when transmissivity dips. As expected, we see that the contour lines are almost the reverse of the transmissivity spectrum in Fig.~\ref{fig:water}(a).

In real application scenarios, there will be additional loss in experimental implementations, which will be improved as the engineering capability advances. Therefore, in this section, we have focused on the fundamental loss from water to provide a performance limit.

\begin{figure}[t]
    \centering
    \includegraphics[width=\linewidth]{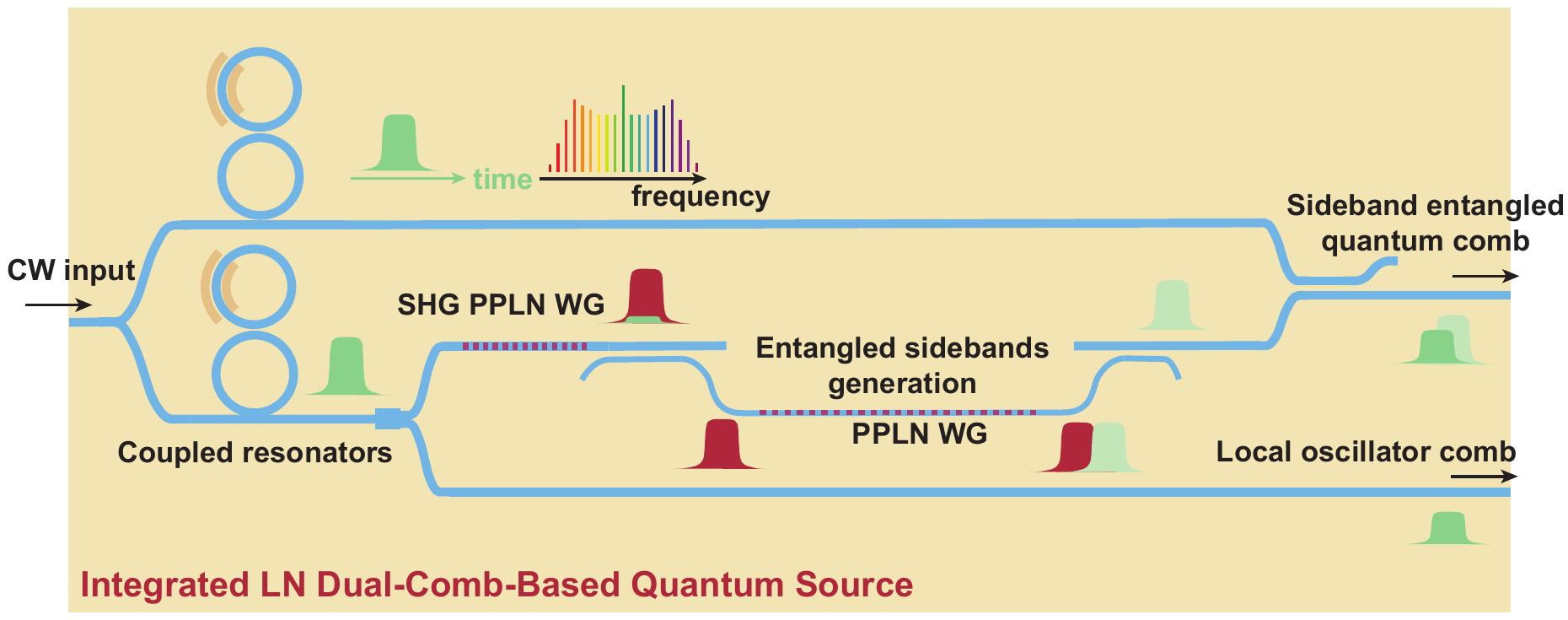}
    \caption{\QZ{Integrated LN photonic circuits for realization of entanglement based dual comb sources. Two classical comb sources with slightly different comb line spacings are generated by pumping a CW laser into coupled microresonators at normal group velocity dispersion regimes. The comb on the bottom side is split into two beam paths: one beam serves as the local comb oscillator while the other beam is sent to a PPLN waveguide for second harmonic generation, after which the SHG beam is used to generate entangled sidebands around each local comb line using a second dispersion engineered PPLN waveguide. The entangled sideband beam is then combined with the classical signal comb, together serving as the quantum signal comb output. The darker green pulse corresponds to the frequency comb at the fundamental frequency while the red pulse corresponds to the frequency comb as second harmonic frequency and the lighter green pulse illustrates the sideband engineered state at the fundamental frequency. SHG, second harmonic generation; PPLN WG, periodically poled lithium niobate waveguide; CW, continuous wave.}}
    \label{fig:exp}
\end{figure}

\section{Discussion}
Before closing, we discuss about experimental generation of the quantum comb. \QZ{We believe that the integrated lithium niobate (LN) photonic platform is well suited for the proposed scheme thanks to its large second order and Kerr nonlinearity as well as capability of achieving quasi-phase matching via electrical poling~\cite{boes2023lithium}. Illustration of the LN photonic chip is shown in Fig.~\ref{fig:exp}. Two frequency combs with slightly different comb line spacings can be generated via pumping a continuous wave (CW) laser into a coupled-microresonator device at normal group velocity dispersion (GVD). Such normal GVD-based frequency combs are reported to have a high conversion efficiency and high comb line spacing of $\sim100$ GHz~\cite{Kim:19}, which is desirable to generate the following quantum frequency comb state. An alternative solution is an electro-optic-modulator based frequency comb and pulse generator~\cite{yu2022integrated}. Among the two combs, one is adopted as the classical signal comb, which will further be combined with entangled sidebands to further generate the quantum signal comb (top side of Fig.~\ref{fig:exp}). The other comb is split into two beams (bottom side of Fig.~\ref{fig:exp}). While one of the beam is stored as the local oscillator comb, the other beam is sent to} a periodically-poled LN (PPLN) waveguide to generate a comb source at the second harmonic frequencies, which is then used to pump another PPLN waveguide for two-mode squeezing around each of the local comb line via spontaneous parametric down-conversion (SPDC), during which the original local frequency comb serves as the seed. By appropriately controlling the phase between the two frequency combs, one can operate in the parametric de-amplification regime to generate amplitude squeezed state at each comb line. The phase matching bandwidth of the SPDC needs to be less than half of the comb line spacing~\cite{zhong2009high}. The sideband entangled outputs are then combined with the \QZ{classical signal} frequency comb \QZ{(shown in the top side of Fig.~\ref{fig:exp})} at a highly transmissive beamsplitter, forming the quantum frequency comb source in the proposed dual-comb configuration. \QZ{As compared to the bulk optical system, the integrated photonic platform can achieve the stringent phase-matching bandwidth at each comb line over a broad optical bandwidth since both the GVD and group velocity can be tailored via engineering of the nanophotonic waveguide dimension. In addition, the high repetition rate of the chip-based Kerr OFC sets a higher upper bound of the phase matching bandwidth. The RIN of Kerr combs can be suppressed with phase-lock-loops and external-cavity acousto-optic or electro-optic actuators \cite{WangOE_RIN_supression}, which can be integrated with the thin film LN platform.}

\QZ{Squeezing and two-mode squeezing have been routinely demonstrated in experiments. For example, 3 dB of squeezing has been demonstrated with bulk optics in radio-frequency sensing squeezing~\cite{xia2020demonstration}. In on-chip system,
1.6 dB of squeezing has recently been demonstrated~\cite{yang2021squeezed}, slightly lower than bulk optics due to extra coupling from chip to fiber. From these results, we expect a near-term demonstration of the proposed quantum dual-comb system to provide 3 dB of quantum advantage, with further efforts in improving device imperfections. The major limitation in the measured squeezing in these systems is the loss from the source to the detection, which can be improved with engineering. For example, with ingenious system design and engineering, Ref.~\cite{vahlbruch2016detection} is able to achieve 15dB of squeezing in a bulk optical platform.}


Note that the proposed quantum comb in this work is different from Ref.~\cite{yang2021squeezed}: there the comb lines are themselves pair-wise entangled to serve as resource for quantum computation; while our proposed quantum comb has side-band of each comb line pair-wise entangled to benefit dual-comb spectroscopy sensing precision.


In this work, we proposed a entanglement-enhanced dual-comb spectroscopy protocol, where both the signal comb and local comb can be quantum engineered. When local comb is stronger than the signal comb, the protocol promises signal-to-noise ratio advantages in detecting low-loss samples such as thin slice of bio-tissues and molecular gas. When the local comb is weak compared with signal comb, quantum engineering of the local comb provides signal-to-noise ratio advantages regardless of the sample loss, making the quantum advantage robust against experimental imperfections.

Dual-comb interferometry is evolving into one of the most powerful tools for broadband laser spectroscopy, ranging and imaging, and our work extends its advantages beyond the standard quantum limit. Such a potential will enable new comb-based spectroscopy and metrology with unprecedented precision and sensitivity. The boost in SNR will directly lead to orders of magnitude improvement in measurement speeds for real-time sensing and imaging.

\appendix

\section{Two-mode squeezed vacuum}
\label{app:TMSV}

In this paper, we are interested in the two-mode squeezed vacuum (TMSV) state, the continuous-variable version of Einstein-Podolsky-Rosen (EPR) state~\cite{weedbrook2012gaussian}. Consider two modes $\hat{a}_1$ and $\hat{a}_2$, with real and imaginary quadrature operators $\hat{q}_j\equiv(\hat a_j+\hat a_j^\dagger)/\sqrt{2}, \hat{p}_j\equiv(\hat a_j-\hat a_j^\dagger)/\sqrt{2}i$, $j=1,2$. The entanglement between the two modes is described by the joint squeezing on modes
$\hat a_+=(\hat a_1+\hat a_2)/\sqrt{2}$ and $\hat a_-=(\hat a_1-\hat a_2)/\sqrt{2}$, such that the variances of joint quadratures $\hat q_+, \hat p_-$ are suppressed to $e^{-2r}/2\equiv 1/2G$. On the other hand, the variances of $\hat q_-, \hat p_+$ are amplified to $e^{2r}/2\equiv G/2$.
In this case, when $r\to \infty$ it has the ideal EPR correlation of $\hat p_1=\hat p_2$, $\hat q_1=-\hat q_2$. 

Take the phase-matched case ($\alpha_n=\beta_n$) of Eq.~\eqref{eq:quadrature_def} and index $\hat a_{n,m}\to \hat{a}_1$ and $\hat a_{n,-m}\to \hat{a}_2$, the operator
$
\hat X_{n,m}\equiv \hat a_1+\hat a_2^\dagger=\hat{q}_++i \hat{p}_-. 
$
Therefore, the TMSV state enables $\text{var}\hat X_{n,m}$ to be suppressed to $e^{-2r}\equiv 1/G$. This is the keystone of the SNR improvement proposed in this paper. For the general case, 
\begin{align}
\hat{X}_{n,m}=&
\cos(\alpha_n-\beta_n)\hat{q}_+
-\sin(\alpha_n-\beta_n)\hat{p}_+
\nonumber
\\
&
+i\cos(\alpha_n-\beta_n)\hat{p}_-
+i\sin(\alpha_n-\beta_n)\hat{q}_-,
\label{X_qp}
\end{align}
from Eq.~\eqref{eq:quadrature_def}. Then, one can obtain the variance in Eq.~\eqref{eq:varX_TMSV}.\\

\section{Derivation of the quantum noise in dual-comb spectroscopy} 
\label{app:derivation}

\QZ{
Here we derive the mean and noise formulas Eqs.~\eqref{eq:Nac_mean} and \eqref{eq:Nac} of the random readout of the AC field $\hat{N}_{\rm AC}=\overline N_{\rm AC}(t)+\hat \Sigma_{\rm AC}(t)$.
}

\QZ{
After the channel described by Eq.~\eqref{eq:IOrelation}, the dual-comb channel inputs Eqs.~\eqref{eq:fullinput_A}\eqref{eq:fullinput_B} become
\begin{widetext}
\bal 
\hat A^\prime(t)&= \frac{1}{\sqrt T}\left[
\sum_{n=1}^N \sum_{{k}=-N}^{N} \hat a_{n,k}^\prime  e^{i 2\pi (n f_r + {k} \Delta f_r) t} +\sum_{n=1}^N
 \sqrt{\kappa_n}A_ne^{i\alpha_{n}} e^{i 2\pi n(f_r+\Delta f_r)t} \right]\equiv \frac{1}{\sqrt T}\sum_{n=1}^N \sum_{{k}=-N}^{N} \hat A^\prime_{n,k}e^{i2\pi (nf_r+k \Delta f_r)t}\,,\\
\hat B^\prime(t)&=\frac{1}{\sqrt T}
\left[\sum_{n=1}^N \sum_{{k}=-N}^{N} \hat b_{n,k}^\prime e^{i 2\pi (n f_r + {k} \Delta f_r) t} + \sum_{n=1}^N  \sqrt{\eta_n}B_ne^{i\beta_{n}} e^{i 2\pi n f_r t}\right]\equiv \frac{1}{\sqrt T}\sum_{n=1}^N \sum_{{k}=-N}^{N}  \hat B^\prime_{n,k}e^{i2\pi (nf_r+k \Delta f_r)t}\,,
\eal
\end{widetext}
where we have defined the frequency components at $f=nf_r+k \Delta f_r$ for the returned signal and LO as
\bal 
\hat A^\prime_{n,k}&= 
\hat a_{n,k}^\prime
 +\delta_{n,k} \sqrt{\kappa_n}A_ne^{i\alpha_{n}}   \,,\\
\hat B^\prime_{n,k}&=
\hat b_{n,k}^\prime
+  \delta_{0,k} \sqrt{\eta_n}B_ne^{i\beta_{n}} \,.
\label{Anm_Bnm}
\eal
Here $\delta_{n,k}=1$ for $n=k$, zero otherwise and the channel output noises $\hat a_{n,k}^\prime, \hat b_{n,k}^\prime$ are defined by Eq.~\eqref{eq:IOrelation}. Note that the noises are zero-mean: $\expval{\hat{a}_{n,k}^\prime}=\expval{\hat{b}_{n,k}^\prime}=0$.
}

\QZ{
As we explained in the main text, the receiver collects the two combs together, interferes the signal comb and the LO comb via a balanced beamsplitter to obtain the fields
$\hat c_\pm(t)$.
One then performs photodetection to obtain the difference current
\bal 
\hat N(t)&=\hat A^{\prime\dagger}(t)\hat B^\prime(t)+\hat B^{\prime\dagger}(t)\hat A^\prime(t)\,,
\eal 
which consists both the mean field and the noise. In the difference current $\hat N(t)$, we obtain a wide range of frequencies, equal to the frequency difference between each frequency component pair in signal and LO including cross-tooth high-frequency terms of frequency $f\sim O(f_r)$. Filtering out the DC term and high-frequency terms of $O(f_r)$, we keep only $f\sim\Delta f_r\ll f_r$ frequency components
\bal 
\hat{N}_{\rm AC}(t)&=
\sum_{n=1}^N  \sum_{k\neq k^\prime} \hat B_{n,k}^{\prime\dagger} \hat A_{n,k^\prime }^{\prime}  e^{i 2\pi (k^\prime-k) \Delta f_r t} +  {\rm h.c.}
\eal
By taking Fourier transform (omitting the delta-function envelops that describes the comb linewidth), we obtain the spectrum at $m\Delta f_r$ for $m\ge 1$,
\be 
\hat{N}_{\rm AC}(m\Delta f_r)=\sum_{n=1}^N \sum_{k=-N}^N \hat B_{n,k}^{\prime\dagger} \hat A_{n,k+m }^{\prime}+\hat B_{n,k}^{\prime} \hat A_{n,k-m }^{\prime\dagger}\,.
\label{NACkf}
\ee
Any negative frequency component of $m\le -1$ is fully determined by its positive frequency component as $\hat{N}_{\rm AC}(t)$ is real and $\hat{N}_{\rm AC}(m\Delta f_r)=\hat{N}_{\rm AC}^\dagger(-m\Delta f_r)$, so it is sufficient to measure the $m\ge 1$ positive frequency components only.
}

\QZ{
For the mean, after plugging in Eq.~\eqref{Anm_Bnm} we have
\begin{align}
&\expval{\hat{N}_{\rm AC}(m\Delta f_r)}
\nonumber
\\
&=\sum_{n=1}^N \sum_{k=-N}^{N} 
\left[
\delta_{0,k} \sqrt{\eta_n}B_ne^{-i\beta_{n}}\delta_{n,k+m} \sqrt{\kappa_n}A_ne^{i\alpha_{n}}
\right.
\nonumber
\\
&
\left.  \qquad \qquad \qquad
+\delta_{0,k} \sqrt{\eta_n}B_ne^{i\beta_{n}}\delta_{n,k-m} \sqrt{\kappa_n}A_ne^{-i\alpha_{n}}
\right]
\\
&
=\sqrt{\eta_m}B_me^{-i\beta_{m}} \sqrt{\kappa_m}A_me^{i\alpha_m},
\end{align}
where the second term of the first line vanishes since we collect only $m\ge 1$ terms. This recovers Eq.~\eqref{eq:Nac_mean}.
}

\QZ{
For the noise, we obtain the leading order contribution by replacing one annihilation operator in the quadratic terms of Eq.~\eqref{NACkf} with its mean (which contains a strong displacement) and the other with its weak noise part, i.e.,
\bal 
~&\hat \Sigma_{\rm AC}(m\Delta f_r)\\
&\simeq \sum_{n=1}^N \sum_{{k}=-N}^{N} \expval{\hat B_{n,k}^{\prime\dagger}} \hat a_{n,k+m }^{\prime}  + \hat b_{n,k}^{\prime\dagger} \expval{\hat A_{n,k+m }^{\prime}}\\
&\qquad \qquad \quad + \expval{\hat B_{n,k}^{\prime}} \hat a_{n,k-m }^{\prime\dagger} + \hat b_{n,k}^{\prime} \expval{\hat A_{n,k-m }^{\prime\dagger}} \\
&=\sum_{n=1}^N \sum_{{k}=-N}^{N}  \delta_{0,k} \sqrt{\eta_n}B_n
\left(e^{-i\beta_{n}}  \hat a_{n,k+m }^{\prime}+e^{i\beta_{n}}  \hat a_{n,k-m }^{\prime\dagger}\right)\\
&\qquad \qquad \quad +    \sqrt{\kappa_n}A_n \left(\delta_{n,k+m} e^{i\alpha_{n}} \hat b_{n,k}^{\prime\dagger}+ \delta_{n,k-m} e^{-i\alpha_{n}} \hat b_{n,k}^{\prime} \right)\\
&=\sum_{n=1}^N    \sqrt{\eta_n}B_n \left(e^{-i\beta_{n}} \hat a_{n,m }^{\prime}+ e^{i\beta_{n}} \hat a_{n,-m }^{\prime\dagger}\right)\\
&\qquad \quad + \sqrt{\kappa_n}A_n \left(e^{i\alpha_{n}}  \hat b_{n,n-m}^{\prime\dagger} + e^{-i\alpha_{n}}  \hat b_{n,n+m}^{\prime} \right)
\label{Noise_AC}
\eal
After plugging the noise input-output relation Eq.~\eqref{eq:IOrelation} in, this recovers Eq.~\eqref{eq:Nac}.
}\\

\section{ Justification of the SNR definition}
\label{app:justification_SNR}

First, we construct the estimator for $\kappa_n$ and show that the estimation error is connected to our definition of SNR in Eq.~\eqref{eq:SNR_bothquad}.
Consider the real quadratures $\hat q_{\rm AC}(m\Delta f_r) \equiv \Re \overline N_{\rm AC}(m\Delta f_r)+\Re \hat \Sigma_{\rm AC}(m\Delta f_r) $, $\hat p_{\rm AC}(m\Delta f_r) \equiv  \Im \overline N_{\rm AC}(m\Delta f_r)+\Im \hat \Sigma_{\rm AC}(m\Delta f_r)  $ of the complex heterodyne readout $N_{\rm AC}$ [see Eq.~\eqref{eq:Nac_mean} and Eq.~\eqref{eq:Nac}]. Their means are given by Eq.~\eqref{eq:Nac_mean} as
\bal 
\expval{\hat q_{\rm AC}(m\Delta f_r) } &= \sqrt{\eta_m \kappa_m }  B_m A_m \cos(\alpha_m-\beta_m) \,,\\
\expval{ \hat p_{\rm AC}(m\Delta f_r) } &= \sqrt{\eta_m \kappa_m }  B_m A_m \sin(\alpha_m-\beta_m)\,.  \\
\eal 
The quadrature fluctuations of Eq.~\eqref{eq:Nac} are sums of contributions from $N$ comb lines
\bal
~&\text{var } \hat q_{\rm AC}(m\Delta f_r)\\
&=\sum_{n=1}^N \left[\frac{\calN_n}{2} +\eta_n \kappa_n \left(B_n^2{\rm var} \Re\hat{X}_{n,m}+A_n^2{\rm var}\Re\hat{Q}_{n,m} \right)\right] \,,\\
&\text{var } \hat p_{\rm AC}(m\Delta f_r)\\
&=\sum_{n=1}^N \left[\frac{\calN_n}{2} +\eta_n \kappa_n \left(B_n^2{\rm var} \Im\hat{X}_{n,m}+A_n^2{\rm var}\Im\hat{Q}_{n,m} \right)\right] \,.
\label{eq:quadnoise}
\eal
The sum of the two noise gives Eq.~\eqref{eq:var_NAC_full}.  From Eq.~\eqref{X_qp} for TMSV state, we can see that $  \Re\hat{X}_{n,m},  \Im\hat{X}_{n,m} ,\Re\hat{Q}_{n,m}, \Im\hat{Q}_{n,m}$ are mutually independent when $\alpha_n=\beta_n$. 

We begin with the estimation of $\kappa_n$'s, assuming perfect phase matching, $\alpha_n=\beta_n$.
In this case, two-mode squeezing has $\text{var } \Re\hat{X}_{n,m}=\text{var } \Im\hat{X}_{n,m}=\frac{1}{2}\text{var } \hat{X}_{n,m}$, $\text{var } \Re\hat{Q}_{n,m}=\text{var } \Im\hat{Q}_{n,m}=\frac{1}{2}\text{var } \hat{Q}_{n,m}$, thus
\be 
\text{var } \hat q_{\rm AC}(m\Delta f_r) =\text{var } \hat p_{\rm AC}(m\Delta f_r) =\frac{1}{2}\text{var } N_{\rm AC}(m\Delta f_r)\,.
\ee 
Also, note that $ \hat q_{\rm AC}$ and $\hat p_{\rm AC}$ commute, and indeed they are mutually independent Gaussian variables [which can be verified from Eq.~\eqref{X_qp}]. Thus, we can define the distribution of the readouts $q, p$ for $\hat q_{\rm AC}(m\Delta f_r)$, $\hat p_{\rm AC}(m\Delta f_r)$ as $P_q(q)\cdot P_p(p)$.
Then the minimum mean square error (MMSE) for unknown parameters $\kappa_n$ is given by the Cram\'{e}r-Rao lower bound of Gaussian distribution:
\begin{align}
&
[\text{MMSE}\sqrt{\tilde\kappa_m}]^{-1}
\nonumber
\\
&\equiv  \left[ \int dq \left(\frac{\partial\log P_q(q)}{\partial \sqrt{\kappa_m}}\right)^2 P_q(q) \right]^{-1}
\\
&\quad +
\left[\int dp \left(\frac{\partial\log P_p(p)}{\partial \sqrt{\kappa_m}}\right)^2 P_p(p)\right]^{-1}\\
&= \frac{\left|\frac{d \overline N_{\rm AC}(m\Delta f_r)  }{d\sqrt{\kappa_m}}\right|^2}  { \text{var } \hat q_{\rm AC}(m\Delta f_r) }+ 2\cdot \frac{\left|\frac{d \text{var } \hat q_{\rm AC}(m\Delta f_r) }{d\sqrt{\kappa_m}}\right|^2}  {2 [\text{var } \hat q_{\rm AC}(m\Delta f_r)]^2 } \\
&\simeq \frac{\left|\frac{d \overline N_{\rm AC}(m\Delta f_r)  }{d\sqrt{\kappa_m}}\right|^2}  { \text{var } \hat q_{\rm AC}(m\Delta f_r) } = \frac{\eta_m B_m^2 A_m^2}  { \text{var } \hat q_{\rm AC}(m\Delta f_r) }\,.
\end{align}
In the last equality, we have assumed that the modulation on the readout variance is negligible, which is true due to squeezing power much lower than the comb power.
Thus, in comparison with Eq.~\eqref{eq:SNR_bothquad},
\be 
[\text{MMSE}\sqrt{\tilde\kappa_m}]^{-1}=2/\kappa_m\cdot \text{SNR}^2\,.
\ee

Now we estimate the phase mismatch $\theta_m\equiv\alpha_m-\beta_m$. Note that $\left|\frac{d \overline N_{\rm AC}(m\Delta f_r)  }{d\theta_m}\right|^2=2\cdot \kappa \eta A^2 B^2$. Similar to the transmissivity estimation above, we assume a TMSV input state such that $\text{var } \hat q_{\rm AC}(m\Delta f_r) =\text{var } \hat p_{\rm AC}(m\Delta f_r)$ and the independence between $\hat q_{\rm AC}$ and $\hat p_{\rm AC}$ still holds. Similarly, the Cram\'{e}r-Rao lower bound gives
\bal 
[\text{MMSE }\theta_m]^{-1} 
= \frac{\eta_m \kappa_m B_m^2 A_m^2}  { \text{var } \hat q_{\rm AC}(m\Delta f_r) } 
\eal
Note that $\text{var } \hat q_{\rm AC}(m\Delta f_r)=\frac{1}{2}\text{var } N_{\rm AC}(m\Delta f_r)$. Thus, in comparison with Eq.~\eqref{eq:SNR_bothquad},
\be 
[\text{MMSE }\theta_m]^{-1}=2\cdot \text{SNR}^2\,.
\ee

\section{Full formula of estimation error}
\label{app:SNR_full}

Here we derive Eq.~\eqref{SNR_overall}, including the inverse-square-law, the inverse-law, and constant noise terms with respect to source power $P_{\rm S}$. We denote the total power of the signal or the LO as $P_{\rm S}$ or $P_{\rm LO}$, and define their ratio $\gamma\equiv  P_{\rm LO}/P_{\rm S}$. We can identify $TP_{\rm S} =h\nu_0 \sum_{m=1}^N |A_m|^2$, $TP_{\rm LO} =h\nu_0 \sum_{m=1}^N |B_m|^2$, where $h\nu_0$ is the energy per photon. To simplify the formulas we assume symmetric comb lines $A_m=A, B_m=B$ for any $1\le m\le N$.
To connect to the SNR of Eq.~\eqref{eq:SNR_bothquad}, we normalize each noise by the power of mean field Eq.~\eqref{eq:Nac_mean}, $\overline N_{\rm AC}(m\Delta f_r)^2=\eta_m \kappa_m A^2 B^2=\eta_m \kappa_m \cdot (P_{\rm S}T/Nh\nu_0)\cdot (P_{\rm LO}T/Nh\nu_0)$.

The inverse-square-law noise comes from the detector noise. It is modelled as a fluctuation of constant noise equivalent power (NEP) on the photon current readout, including dark
current, Johnson noise, amplifier noise figure, etc.. Note that NEP-type noise is circular symmetric in the phase space, it affects both real and imaginary parts of $N_{\rm AC}$. Since we defined $\text{var } N_{\rm AC}$ as the sum of the two quadrature noises, the physical NEP-type noise is $2\text{NEP}^2\Delta f$ in power. In photon number, the noise is $2(\text{NEP}\cdot T/Nh\nu_0)^2\Delta f$. Here the bandwidth is defined as $\Delta f=1/2T$ for single-sided NEP spectral density. According to Eq.~\eqref{eq:SNR_bothquad}, the detector noise results in the normalized NEP-type noise at intermediate frequency $m\Delta f_r$
\be 
\sigma^2_{ \rm NEP}=2\frac{(\text{NEP}\cdot T/Nh\nu_0)^2\Delta f}{\eta_m \kappa_m (P_{\rm LO}T/Nh\nu_0) (P_{\rm S}T/Nh\nu_0)}=\frac{N^2}{T}\frac{\text{NEP}^2}{ \eta_m \kappa_m \gamma P_{\rm S}^2}\,.
\ee
Here NEP is has the unit of W/Hz$^{1/2}$. Now we see that the NEP-type noise-power relation is $\sigma^2_{ \rm NEP}\propto\frac{1}{P_{\rm S}\cdot P_{\rm LO}}$, which is an inverse-square-law term $\sim O(\frac{1}{P_{\rm S}^2})$ when $\gamma$ is fixed.

The inverse-law noise comes from the intrinsic quantum noise Eq.~\eqref{eq:Nac}. For the case where phase noise is negligible, the quadrature fluctuation leads to the normalized quadrature noise
\be 
\sigma^2_{ \rm quad}=  \frac{ \text{var } \hat \Sigma_{\rm AC}}{\eta_m \kappa_m B^2 (P_{\rm S}T/Nh\nu_0)}=\frac{N^2}{T}c_\gamma\frac{4h\nu_0 }{ P_{\rm S}},
\ee
where 
\be 
c_\gamma\equiv\frac{\text{var } \hat \Sigma_{\rm AC}}{4 \eta_m \kappa_m B^2\cdot N }\ee  
with $\text{var } \hat \Sigma_{\rm AC}$ defined in Eq.~\eqref{eq:var_NAC_full}. For vacuum input $G=1,G_{\rm LO}=1$, unit transmissivities $\eta_n=\kappa_n=1$ and zero noise $\tilde{\calE}_n=0$ for any $1\le n\le N$, we have $c_\gamma=\frac{1}{4}(1+\frac{1}{\gamma})$. In this case, $\text{var } \hat \Sigma_{\rm AC}\propto P_{\rm S}+P_{\rm LO}$, we see that the fundamental noise-power relation is $\sigma^2_{ \rm quad}\propto\frac{P_{\rm S}+P_{\rm LO}}{P_{\rm S}\cdot P_{\rm LO}}$, which is an inverse-law term $\sim O(1/P_{\rm S})$ when $\gamma$ is fixed. 

The constant noise results from the relative intensity noise (RIN). Consider the power of each comb tooth, $ h\nu_0 A^2/T$ for signal and $ h\nu_0 B^2/T$ for LO. RIN is modelled as a single-sided white noise $\text{var } (h\nu_0 A^2/T)=(\text{RIN} \Delta f) P_{\rm S}^2$, $\text{var }(h\nu_0 B^2/T)=(\text{RIN} \Delta f) P_{\rm LO}^2$ on the power of field amplitudes $A, B$, generated from additive amplified spontaneous emission (ASE) from the laser or from any subsequent optical amplification. Similar to NEP-type noise, the bandwidth is defined as $\Delta f=1/2T$ for single-sided RIN spectral density. RIN-type noise is also circular symmetric in the phase space, which affects both real and imaginary parts of $ N_{\rm AC}$ and yields a factor of 2 in the complex-observable variance $\text{var } N_{\rm AC}$. Consequentially, the physical noise is
\bal 
~&2 \eta_m \kappa_m [\text{var }(A)\cdot  B^2+  A^2\cdot \text{var }(B)]\\
&=2\eta_m \kappa_m  [\frac{\text{var }(A^2)}{4P_{\rm S}}\cdot P_{\rm LO}+  P_{\rm S}\cdot \frac{\text{var }(B^2)}{4P_{\rm LO}}]\\
&=2\eta_m \kappa_m \left(\frac{T}{h\nu_0}\right)^2\cdot \text{RIN} \Delta f \cdot  P_{\rm S}^2 (\frac{1}{4}\cdot  \gamma +   \frac{\gamma^2}{4\gamma})\\
&=\eta_m \kappa_m \left(\frac{T}{h\nu_0}\right)^2\cdot \text{RIN} \Delta f \cdot \gamma P_{\rm S}^2
\label{eq:physicalRIN_app}
\eal 
Here RIN is in unit 1/Hz, and in the first equality we have used $\text{var }(A^2)=(\frac{\partial}{\partial A}A^2)^2\text{var }(A)=4A^2\text{var }(A)$, $\text{var }(B^2)=4B^2\text{var }(B)$, assuming $A^2\gg \text{var }A$ and $B^2\gg \text{var }B$. It is valid to consider fluctuations on amplitudes $A,B$ instead, because that the RIN physically comes from the amplified spontaneous emission, which is modelled as a Gaussian noise on the field quadratures in quantum optics.
The amplitude fluctuations result in the normalized RIN-type noise
\be 
\sigma^2_{ \rm RIN}=\frac{\eta_m \kappa_m \text{RIN} \Delta f \cdot \gamma (P_{\rm S}T/h\nu_0)^2}{\eta_m \kappa_m (P_{\rm LO}T/Nh\nu_0) (P_{\rm S}T/Nh\nu_0) }=\frac{N^2}{T}  2c_{\gamma^2}   \text{RIN} \,.
\ee
Here the coefficient $c_{\gamma^2}\equiv 1/4$.
We immediately see that the RIN-type noise-power relation is $\sigma^2_{ \rm RIN}\sim O(1)$, which does not depend on the signal or LO power.

Overall, we can write the full formula of the SNR Eq.~\eqref{eq:SNR_bothquad} at intermediate frequency $m\Delta f_r$ as $\text{SNR}^{-2}= \sigma^2_{ \rm NEP}+\sigma^2_{ \rm quad}+\sigma^2_{ \rm RIN}$, which gives Eq.~\eqref{SNR_overall} in main text.

To recover the classical results in Ref.~\cite{Newbury:10}, we further assume that the frequency spectra of all parameters are almost uniform, such that $\kappa_m\approx \kappa, \eta_m\approx \eta$ for any $1\le m\le N$. When $\eta, \kappa\to 1$ and classical source is used ($G=G_{\rm LO}=1$), our result recovers the formula of $\sigma_H$ [Eq.~(2)] in Ref.~\cite{Newbury:10} mostly [$H(f)$ is the transfer function of electrical field, equivalent to $\sqrt{\kappa(f)}$]. Note that in the quantum model the shot noise term $a_{\rm shot}$ is instead formulated as the quadrature fluctuation $a_{\rm quad}$, while the quadrature fluctuation incidentally gives a similar result $c_{\gamma}=\frac{1}{4}(1+\frac{1}{\gamma})$, up to an extra $1/2$ factor. Also, $c_{\gamma^2}=\frac{1}{4}\cdot \frac{2\gamma}{2\gamma}$ is different from $(1+\gamma^2)/2\gamma$ in Eq.~(2) of Ref.~\cite{Newbury:10}, which is the result of our derivation of the physical RIN-type noise Eq.~\eqref{eq:physicalRIN_app} in the balanced detection.
The independence of $c_{\gamma^2}$ and thereby of $\sigma^2_{\rm RIN}$ in our Eq.~\eqref{SNR_overall} agrees with the recent results of Eqs.~(58)(59) in Ref.~\cite{wissel2022relative} assuming white noise spectrum. 
Meanwhile, our derivation recovers the result $c_{\gamma^2}^{\rm unbal}=\frac{1+\gamma^2}{2\gamma}$ in Ref.~\cite{Newbury:10} for the unbalanced detection case where the physical noise is $\sim \text{var }(A^2)+\text{var }(B^2)$ instead. To summarize, our result when applied to the classical dual-comb SNR can be obtained by letting $c_{\gamma}=\frac{1}{4}(1+\frac{1}{\gamma})$, $b=1$, $c_{\gamma^2}=1/4$ in Eq.~(2) of Ref.~\cite{Newbury:10}.

We note that our quantum model yields an SNR-gamma relation different from the semiclassical model in Ref.~\cite{Newbury:10}. For example, at a fixed $P_{\rm S}$ there is a finite optimum value of $\gamma$ to maximize the SNR in the semiclassical model while the optimum is at $\gamma\to \infty$ in our quantum model. This is because when $\gamma$ is large, the RIN-type noise increases with $\gamma$ in the semiclassical model, while RIN-type noise remains a constant in our derivation which agrees with Ref.~\cite{wissel2022relative}.

Finally, we address the saturation in the SNR with respect to the squeezing gain $G$, due to NEP or RIN noise. For simplicity, we consider $\kappa_m\approx \kappa, \eta_m\approx \eta$ for any $1\le m\le N$. The saturation of SNR begins when the squeezing gain $G$ is large enough such that NEP-type or RIN-type noise overwhelms the fundamental noise. By solving $\sigma^2_{ \rm NEP}>\sigma^2_{ \rm quad}$ or $\sigma^2_{ \rm RIN}>\sigma^2_{ \rm quad}$, we derive the saturation threshold as
$
P_{\rm S,sat}^{\rm NEP}={G\cdot  \text{NEP}^2}/\{h\nu_0  (\gamma  [G (1-\kappa)+\kappa]+\kappa)\}
$
for the NEP-type and 
$
P_{\rm S,sat}^{\rm RIN}={h\nu_0  (\gamma  (G (1-\kappa)+\kappa)+\kappa)}/\{G \gamma  \kappa \cdot  \text{RIN}\} 
$ 
for the RIN-type. The above thresholds are indicated in Fig.~\ref{fig:overview} as the dots on the SNR curves, showing a good agreement with numerical results.

\begin{acknowledgements}

This work is supported by the National Science Foundation CAREER Award CCF-2142882, Office of Naval Research Grant No. N00014-23-1-2296, National Science Foundation Engineering Research Center for Quantum Networks Grant No. 1941583 and Cisco Systems, Inc.. Z.Z. also acknowledges NSF grant No. ECCS-1920742 and CAREER Award No. ECCS-2144057.

H.S., Z.C., M.J., Z.Z. and Q.Z. proposed the idea of entanglement-enhanced dual-comb spectroscopy protocol in discussions. 
Q.Z. conceived the general methodology for the investigation and supervised the project. 
H.S. performed the analyses and generated the figures, with the exception of Fig. 1b generated by Z.C.. 
Z.C. contributed to the modeling of classical dual-comb spectroscopy, including practical noise and other experimental considerations. M. Y. and Z. Z. contributed to the experimental design of quantum comb generation. S.E.F. contributed to the biological application scenarios.
All authors contributed to the  writing of the manuscript.
\end{acknowledgements}

%

\end{document}